\documentclass[iop]{emulateapj}

\usepackage{natbib}
\usepackage{amssymb}		
\usepackage{subfigure}

\newcommand{\scinot}[2]{ \ensuremath{#1\times10^{#2}} }

\newcommand{\kms}{~km\,s$^{-1}$}
\newcommand{\msol}{~M$_{\sun}$}

\newcommand{\flux}{~erg\,s$^{-1}$\,cm$^{-2}$}
\newcommand{\surfb}{~erg\,s$^{-1}$\,cm$^{-2}$\,arcsec$^{-2}$}
\newcommand{\lum}{~erg\,s$^{-1}$}

\newcommand{\pcc}{~cm$^{-3}$}

\shorttitle{IFU Observations of NGC~4696}
\shortauthors{Farage et al.}

\begin{document}

\title{Optical IFU Observations of the Brightest Cluster Galaxy NGC~4696: \\The Case for a Minor Merger and Shock-excited Filaments}
\author{C. L. Farage, P. J. McGregor,  M. A. Dopita, and G. V. Bicknell}
\affil{Research School of Astronomy \& Astrophysics, Australian National University, Cotter Road, Weston, ACT, 2611, Australia}
\email{cfarage@mso.anu.edu.au; peter@mso.anu.edu.au; mad@mso.anu.edu.au; geoff@mso.anu.edu.au}

% Catherine Farage, cfarage@mso.anu.edu.au, +612 6125 0825
% Peter McGregor, peter@mso.anu.edu.au, +612 6125 8033
% Michael Dopita, mad@mso.anu.edu.au, +612 6284 3509
% Geoffrey Bicknell, geoff@mso.anu.edu.au, +612 6125 9088
% Postal address (all authors): Research School of Astronomy & Astrophysics, Mount Stromlo Observatory, Cotter Road, Weston ACT 2611, Australia.
% Fax: +612 6125 0233

\begin{abstract}
We present deep optical integral-field spectroscopic observations of the nearby ($z\sim0.01$) brightest cluster galaxy NGC~4696 in the core of the Centaurus Cluster, made with the Wide Field Spectrograph (WiFeS) on the ANU 2.3m telescope at Siding Spring Observatory. We investigate the morphology, kinematics, and excitation of the emission-line filaments and discuss these in the context of a model of a minor merger. We suggest that the emission-line filaments in this object have their origin in the accretion of a gas-rich galaxy and that they are excited by $v \sim 100-200$\kms\, shocks driven into the cool filament gas by the ram pressure of the transonic passage of the merging system through the hot halo gas of NGC~4696.\\
\end{abstract}


%--------------------------------------------------------------------
\section{Introduction}\label{sec:intro}

Many of the giant elliptical brightest cluster galaxies (BCGs) in the cores of massive clusters are surrounded by systems of filaments that produce a distinctive LINER\footnote{~Low-ionization nuclear emission-line region \citep{Heckman:1980p10009}}-like emission spectrum.
These structures exist in `cool-core' clusters, which have centrally-peaked X-ray surface brightnesses. The densities and pressures are sufficiently high that the radiative-cooling timescales are less than a Hubble time in the core. As a result, there should be large-scale condensation and inflow of cooling intra-cluster gas into the cluster center \citep{Cowie:1977p10039,Fabian:1977p8925,Fabian:1984p5295}. However, recent observations show that much of the cooling of the cluster gas is being offset by heating associated with the central galaxy \citep[e.g., see][]{Peterson:2006p331}.  

In the current picture, the intra-cluster medium (ICM) cooling is regulated by intermittent energy feedback from the central galaxies. Potential heat sources include the mechanical energy of expanding radio lobes \citep[e.g.][and references therein]{McNamara:2007p345}, conduction from the cluster halo \citep[e.g][]{Voigt:2004p6445}, hot stars \citep{Voit:1997p2496}, and high-energy particles \citep[e.g.][]{Rephaeli:1995p11309}.  Cool-core clusters allow detailed studies of galaxy and AGN feedback at relatively low redshift. They offer insight into analogous processes that occurred during galaxy growth in the high-redshift universe and into the physics needed to incorporate physical descriptions of feedback in galaxy formation models.

The extensive, filamentary emission-line nebulae in cool-core cluster BCGs are likely to be products of the ongoing and inter-related processes of heating and cooling.  However, many questions about the origin of the material in the filaments and the mechanisms that excite the emission spectrum remain largely unanswered.
  
Various mechanisms have been proposed as sources of excitation in the filaments, including photoionization by radiation from the AGN, cluster X-rays, or hot stars (young or old); collisional heating by high-energy particles, shocks or cloud-cloud collisions, and conduction of heat from the X-ray corona \cite[see e.g.][and references therein]{Ferland:2009p7851,Crawford:1992p3755,Donahue:2000p3527,Hatch:2007p2057,Johnstone:1988p4157,Johnstone:2007p2172,Sabra:2000p2757,Wilman:2002p4846}. However, none so far satisfactorily describes the characteristic spectra, energetics, and kinematics of the extended emission-line regions.

Multi-wavelength observations of BCGs reveal that material in a range of phases co-exists in the filaments. Emission detected from various galaxies indicates the presence of cold CO at 10-100\,K, mid-infrared-emitting H$_{2}$ at 300-400\,K, molecular gas at 2000-3000\,K producing near-infrared H$_{2}$ rovibrational lines, atomic gas at $10^{4}$\,K emitting atomic forbidden and hydrogen recombination lines, and a hot atomic component traced by X-ray emission at $10^{6}-10^{7}$\,K \citep[see e.g.][]{Johnstone:2007p2172}. Additionally, these cooler gas phases are embedded in the hot, X-ray-emitting ICM at $\sim10^{7}-10^{8}$\,K. Dust is also intermixed with the emission-line filaments in many systems \citep[e.g.][]{Donahue:1993p11449,Sparks:1989p2171,Hatch:2007p2057}. 

There are also systematic differences in the properties of the nebulae amongst clusters. Surveys show that BCGs with more luminous line emission have: shorter cluster cooling times \citep{ODea:2008p8568}; increased blue/UV excess emission \citep{Heckman:1989p746,Allen:1992p1098,Crawford:2005p334}; greater IR luminosities \citep{Egami:2006p3015,ODea:2008p8568,Quillen:2008p3760,Donahue:2007p7066}; larger masses of molecular gas associated with the filaments and galaxy \citep[e.g.][]{ODea:2008p8568}; and more extensive emission-line regions \citep{Crawford:1999p1097}.
Interestingly, there is also a clear trend of decreasing ratios of hydrogen recombination to collisionally-excited forbidden line flux ([\ion{N}{2}], [\ion{S}{2}]) with increasing total emission-line luminosity of the nebulae \citep{Crawford:1999p1097}. 
Strong IR and blue continuum excesses imply that significant star formation occurs in some BCGs. However, in many cases the emission spectrum does not resemble that from a purely star-forming \ion{H}{2} region.

These results have led some authors to conclude that several excitation mechanisms may participate, and contribute to varying degrees in different systems.
\citet{Voit:1997p2496} suggested that sources of `supplementary heating' produce the LINER-like properties of the spectra, though not necessarily through the same mechanism in all systems. 
Similarly, \citet{Hatch:2006p1101} propose that photoionization by hot young stars becomes increasingly dominant in high-luminosity systems over an underlying excitation source that produces the low-ionization (LINER-like) spectrum and dominates the lower-luminosity nebulae.  
The nature of LINER-producing component(s) is uncertain, however.

We present here a detailed study of the extended filaments surrounding the giant elliptical galaxy \object{NGC~4696}, the brightest cluster galaxy of the Centaurus cluster (Abell 3526). This is a relatively nearby system ($z\sim0.01$), in a cool-core cluster with peaked X-ray surface brightness in the center \citep{Mitchell:1975p8838}. From a pure cooling-flow model the inferred mass deposition rate is $\sim$20\msol\,yr$^{-1}$ \citep{Fabian:1982p2130}.

Extended line emission around NGC~4696 was identified by \citet{Fabian:1982p2130}, and the structures have been imaged in detail by \citet{Crawford:2005p334}. Their narrow-band H$\alpha$ images reveal a complex system of filaments that extends over 50\arcsec\,($\sim$10\,kpc). The galaxy is a prototypical cool-core BCG in which the filaments can be spatially resolved and studied in some detail. 

The H$\alpha$ luminosity of the galaxy emission places it at the low to intermediate end of the range of BCG line luminosity amonst the ROSAT Brightest Cluster Sample compiled by \citet{Allen:1992p1098}. This sample spans four decades of H$\alpha$ luminosity.  The spectral properties are typical of lower-luminosity BCGs, exhibiting a very high ratio of [\ion{N}{2}] to hydrogen recombination emission \citep{Johnstone:1987p4824,Lewis:2003p2494,Sparks:1989p2171}. Since in this case the low-ionization (LINER-like) properties dominate, we can investigate the nature of the LINER excitation with less contamination from other sources of excitation. 

The host galaxy is a giant elliptical central cluster galaxy, but an extended cD stellar halo has not been detected \citep{ArnalteMur:2006p3611,Jerjen:1997p4440}. The surface brightness profile is peaked within the central few arcseconds and there is evidence from the HST I-band image of dual nuclei within the central arcsecond of the core \citep[$\sim$0.2\,kpc;][]{Laine:2003p8500}.
The measured central velocity dispersion (257\kms) implies a central black hole mass on the order of \scinot{(4\pm1)}{8}\msol\, \citep{Rafferty:2006p2278} using the Magorrian $M_{BH}-\sigma$ relation \citep{Magorrian:1998p5262,Gebhardt:2000p3251}.
NGC~4696 is associated with the low-power FR I radio source PKS~1246-41. The irregular lobes of radio emission of this source appear to have been highly distorted by the surrounding medium \citep{Taylor:2006p2047,Taylor:2007p344}. 

Strong dust features are seen in the galaxy, including a conspicuous dust spiral \citep{Shobbrook:1966p8710} that can be traced from the outer regions of the galaxy and in to the galaxy core \citep{Laine:2003p8500}. Close correspondence between the optical emission and dust features is observed in a number of BCGs, and this is well-illustrated in NGC~4696 where dust lanes are associated with most of the extended optical filaments \citep{Sparks:1989p2171,Crawford:2005p334}. The dust and line emission are concentrated to the south of the galaxy nucleus. The dust lane obscures the galaxy light, suggesting that it is positioned on the near side of the galaxy. That a large amount of dust exists in the filaments is a challenge to models in which the cool gas condenses from the hot ICM, because dust is quickly, relative to the cooling timescale, destroyed by sputtering in the hot cluster halo environment \citep{Voit:1995p10873,Donahue:2000p3527}.

\citet{Crawford:2005p334} compare imaging of the extended emission of NGC~4696 in the radio, optical, and X-ray wavebands. The complex system of optical line-emitting filaments is closely traced by soft (0.3 - 1 keV) X-ray emission. The striking resemblance between optical emission from warm gas at $T\sim10^{4}$\,K and X-rays from material at $\sim10^{7}$\,K is also seen in other BCGs, e.g.  M87 and NGC 1275 \citep{Sparks:2004p4125,Fabian:2003p4048}, and is another challenge for models of the excitation. 
The images of \citet{Crawford:2005p334} also illustrate that there is no clear correlation between the morphology of the radio emission and the optical filaments in this system. However, as observed in many cool-core clusters, the lobes of radio emission coincide with cavities in the high-energy X-ray emission from the ICM \citep{Fabian:2005p2036}. 

In this paper we present the kinematics and excitation of the optical line emission detected in the filaments over the central 20\arcsec\, ($\sim4$\,kpc).  Our data show material that is spiralling inward to the galaxy nucleus and uniform excitation properties across the emission regions. The results support the proposal made by \citet{Sparks:1989p2171} that the material in the emission-line filaments surrounding this galaxy originated in a neighboring dwarf galaxy that has undergone a recent merger with the BCG. We describe a scenario in which the line emission is excited by low-velocity shocks that are driven by the motion of the infalling material. We present shock models that reproduce the emission-line flux ratios observed in the gas and surmise that a minor-merger is energetically capable of exciting the observed emission.

Throughout this paper we adopt the cosmological parameters $H_{0} = 71$\,km\,s$^{-1}$\,Mpc$^{-1}$, $\Omega_{M} = 0.27$ and $\Omega_{\Lambda} = 0.73$, based on the five-year WMAP results \citep{Hinshaw:2009p10650}. NGC~4696 has a redshift of $z=0.0099$ \citep{deVaucouleurs:1991p9528} and the spatial scale is 0.21\,kpc\,arcsec$^{-1}$. We assume a heliocentric distance to the galaxy of 44\,Mpc.

%------------------------------------------------------------------------------------------------------------------------------

\section{Observations}\label{sec:obs}
The observations of NGC~4696 were obtained with the Wide-Field Spectrograph \citep[WiFeS;][]{Dopita:2007p6557,Dopita:2010p10019} and the Australian National University (ANU) 2.3m telescope at Siding Spring Observatory. WiFeS is a double-beam, image-slicing, integral-field spectrograph that records optical spectra over a contiguous $25\times38$\arcsec~field-of-view. This spatial field is divided into twenty-five 1\,\arcsec-wide `slitlets', with 0.5\arcsec\,sampling along the 38\arcsec~length. The position angle of the slitlets on the sky was 305\degr~during the observations of NGC~4696.

The WiFeS instrument is a dual-channel device with separate gratings and cameras for the blue and red channels. Each camera is equipped with 4k$\times$4k pixel CCD detectors. In the low-resolution grating configuration that was used for these observations (employing the R3000 and B3000 gratings and the RT560 dichroic), this provides a continuous wavelength coverage from 3290 to 9330\,\AA~in the reduced data. The spectral sampling is 0.81\,\AA/pixel in the blue data cube and 1.31\,\AA/pixel in the red data cube. The average spectral resolution achieved was $\sim$100\kms~FWHM.

The observations of NGC~4696 were made on 14-17 April and 1-3 May 2009.  The nights were clear and the seeing ranged between approximately 1\arcsec\,and 3\arcsec. Multiple 30 minute exposures were obtained for each data set, for total on-source integration times of 14 hours with the blue camera and 18.5 hours with the red (the blue detector was not available during part of the observing period).  
The pattern of science observations consisted of two pairs of galaxy frames separated by a 30 minute exposure of the sky. The scaled sky frame was subtracted from each of the four galaxy frames in this sequence. A dither pattern of 2\arcsec\,offsets was applied to the position of each galaxy frame. A nearby reference star was used to autoguide the galaxy exposures.

Each set of four galaxy frames and one sky frame was preceded and followed by an observation of a spectrophotometric standard star (LTT4364, HD128279, or EG131) and these were used to calibrate the absolute flux scale of the observations. Dwarf stars with spectra that resemble a featureless blackbody (including LTT4364 and EG131) were also measured during the night and used to remove atmospheric absorption features from the flux standard star and galaxy spectra. 
Arc calibration lamp frames and quartz lamp flat-field frames were observed at the beginning of each night, and twilight sky flat-field frames were obtained once during each of the two observing periods.

%--------------------------------------------------------------
\subsection{Data reduction}
The data were reduced using the WiFeS IRAF data reduction package \citep{Dopita:2010p10019}, which is based on and utilizes the Gemini IRAF package\footnote{\url{http://www.gemini.edu/sciops/data-and-results/processing-software}}.
The procedure began by treating each frame as a 2D image. The detector bias signals were removed in the standard way, using the overscan regions of the raw frames. Then the sky frame was scaled and subtracted from the bias-subtracted object frame. 

The spectrograph optically divides the field-of-view into slices along the long axis of the field. These `slitlets' are stacked end-to-end and dispersed across the detector, to form a series of 2D slit spectra from adjacent slices on the sky. In processing, the 2D spectrum from each slice was extracted from the bias- and sky-subtracted frame and placed in an individual FITS image extension. Standard methods for processing 2D spectra were then applied individually within each slice as described below. 

The flat-field frame is formed from exposures of a quartz-iodine lamp.  Within each slice, the spectra were normalized, using a spline function fit to the spectral profiles. The shape of the lamp illumination in the spatial direction along the slices was corrected using twilight sky frames. An additional adjustment was made to maintain the relative signal levels between the slices. The final flat-field frame then retained only pixel-to-pixel variations in the spectral direction and the slit transmission profile in the spatial direction. These effects were subsequently removed from the science data by flat-fielding. Each slice of the galaxy frames was corrected with the corresponding 2D slice from the flat-field frame.

The individual 2D spectra were then transformed to a common spatial and spectral coordinate grid.
NeAr arc lamp frames were used to calibrate the spectral scales of the slices. 
The IRAF \textsc{identify} task was used to match the emission features in the arc lamp spectra, and \textsc{fitcoords} was used to apply the spectral coordinate solution to the science data, slice by slice. The average RMS error in the fit to the arc-line positions is 0.2\,\AA\,for the red cube and 0.1\,\AA\,for the blue cube.
For the spatial alignment, an occulting wire is available at an intermediate focus within the instrument. This provides a convenient spatial registration when the aperture is illuminated by the flat-field lamp. An image of the wire is seen near the center of each slitlet and this establishes the origin of the spatial coordinates in each slice. We assume a standard spatial scale of 0.5\arcsec\,per pixel along the slices.
The data were resampled onto a common rectilinear grid by the \textsc{transform} task and the 2D slices were then stacked together to form a 3D data cube.
%%% referee (2):
The ends of the spectra were trimmed from the reduced data cubes to remove the regions of low sensitivity in the grating+dichroic transmission profile. Approximately 1050\,\AA\, and 550\,\AA\, were removed from the short wavelength end of the red and blue cubes, respectively, and approximately 420\,\AA\, and 160\,\AA\, from the long wavelength end of the cubes. This still provided $\sim500\,$\AA\, of overlap between the cubes. The trimmed regions do not extend to within 400\,\AA\, of the emission lines measured.

Telluric absorption features were corrected in the red galaxy spectra using an integrated spectrum from a reduced smooth-spectrum standard star data cube. We normalized the spectrum to remove the continuum and obtain a spectrum containing only atmospheric absorption features. This was compared to the absorption features seen in the galaxy spectra, scaled and shifted as required, and used to correct the galaxy data using the IRAF \textsc{telluric} task.
After removing the telluric features from a standard star data cube, an integrated flux-standard spectrum was extracted and the standard IRAF flux-calibration tasks used to calibrate the spectra in the galaxy data cubes. The spectra across the whole field are corrected for the shape of the spectral transmission profile as measured at the position of the standard star (approximately the center of the field). The flux calibration was performed without correcting for atmospheric extinction in the individual data frames. The variation in continuum flux among the frames gives an indication of the uncertainties in the flux calibration: there is a 1$\sigma$ variation of $\sim 11$\%, attributable to the changing airmass (the galaxy altitude was between 1 to 1.4 airmasses during the observations) and fluctuations in sky transparency. 

%%% referee (3):
Each science frame was individually reduced by using the procedures described above. The dithered galaxy data cubes were then spatially registered using the continuum centroid of the galaxy nucleus. This was measured for both the red and blue cubes, using an image formed by summing several spectral planes in a line-free region near the center of the spectrum.
The cubes were then offset as required and median-combined with the IRAF task \textsc{imcombine}. Cosmic rays and detector defects were removed in this process.

%-------------------------------------------------------------------------
\subsection{Continuum subtraction}\label{sec:contsub}

Measuring the flux of the extended line-emission regions surrounding NGC~4696 is complicated by the strong stellar continuum. This dominates the spectrum in most regions and must be removed to extract accurate emission-line fluxes. This can be done using the galaxy itself, rather than attempting to generate a theoretical stellar continuum template. 

Since the distribution of line emission around NGC~4696 is asymmetric (see Section~\ref{sec:morph}) and the filaments do not extend beyond approximately 5 to 10\arcsec\, from the galaxy center to the north and east of the nucleus, we were able to extract a spectrum from a region of the inner galaxy that is free from line emission and use this to subtract the stellar component from the spectra at other locations. The representative galaxy spectrum is constructed from a 3\arcsec-diameter aperture placed $\sim8$\arcsec~to the north-east of the continuum center, where no line emission is detected in either the red or blue data cubes, as close to the galaxy core as possible. For each of the strong spectral emission features of interest, this spectrum was scaled to match the flux level of the continuum in neighboring spectral regions (using the average level in two bands approximately 35\,\AA~wide bracketing the line or complex of lines and centered within 50\,\AA~of the feature), before it was subtracted at each spatial pixel. 
%%% referee (4), (5)  -- estimate of uncertainties in continuum subtraction
An estimate of the error introduced in this process is made in the following section.

%-------------------------------------------------------------------------
\subsection{Line profile fits}\label{sec:linefits}

After subtracting the galaxy stellar continuum, we extracted the flux of the emission lines that are detected at better than the 3$\sigma$ level. This corresponds to a line surface brightness $\gtrsim\scinot{4}{-17}$\surfb\, in the red data cube and $\gtrsim\scinot{6}{-17}$\surfb\, in the blue for a line width of 300\kms. The following lines are detected and fit: [\ion{O}{2}] $\lambda\lambda$3726,29\,\AA\, (spectrally unresolved),  [\ion{O}{3}] $\lambda\lambda$4959,5007\,\AA, [\ion{N}{1}] $\lambda\lambda$5198,200\,\AA\, (spectrally unresolved), [\ion{O}{1}] 6300\,\AA, [\ion{N}{2}] $\lambda\lambda$6548,83\,\AA, [\ion{S}{2}]\,$\lambda\lambda$6716,31\,\AA\, as well as Balmer hydrogen recombination lines H$\alpha$, H$\beta$ and H$\gamma$.

We fit single Gaussian line profiles to the emission lines of interest in the continuum-subtracted spectra using the IRAF task \textsc{ngaussfit} and measured line fluxes, the centroid positions, and widths at each spatial pixel having sufficient signal to allow this to be done.
H$\alpha$ and the lines of the [\ion{N}{2}]\,$\lambda\lambda$6548,83\,\AA\,doublet were fit together as three Gaussian components, and the lines of each of the [\ion{S}{2}]\,$\lambda\lambda$6716,31\,\AA\,and [\ion{O}{1}]\,$\lambda\lambda$6300,63\,\AA\,doublets were also fit simultaneously. All the other detected lines were fit individually. 

Radial velocities are given relative to the velocity of the galaxy with respect to the local standard of rest. The velocities have been corrected for the motion of the earth with respect to the local standard of rest; the magnitude of the correction required was less than 9.5\kms\,for all observations. 
The instrumental resolution has been removed from the fitted line widths by subtracting in quadrature the average value of 97\kms~FWHM that was measured from an arc lamp calibration image.
The results derived from the data cube and the emission-line fits are discussed in the following sections. 

%%% referee (6)
Uncertainties in the relative line flux measurements were estimated using the noise properties of the spectra after continuum subtraction. For a given emission line, the $1\sigma$ error in the flux density per pixel was taken to be the standard deviation in a line-free region adjacent to the feature. These were combined in quadrature to form the uncertainties in the integrated line flux. 

%-------------------------------------------------------------------------------------------------------------------------------

\section{Results} \label{sec:results}

\subsection{Galaxy continuum}\label{sec:galcont}
\begin{figure*}
\epsscale{1.14}
\plotone{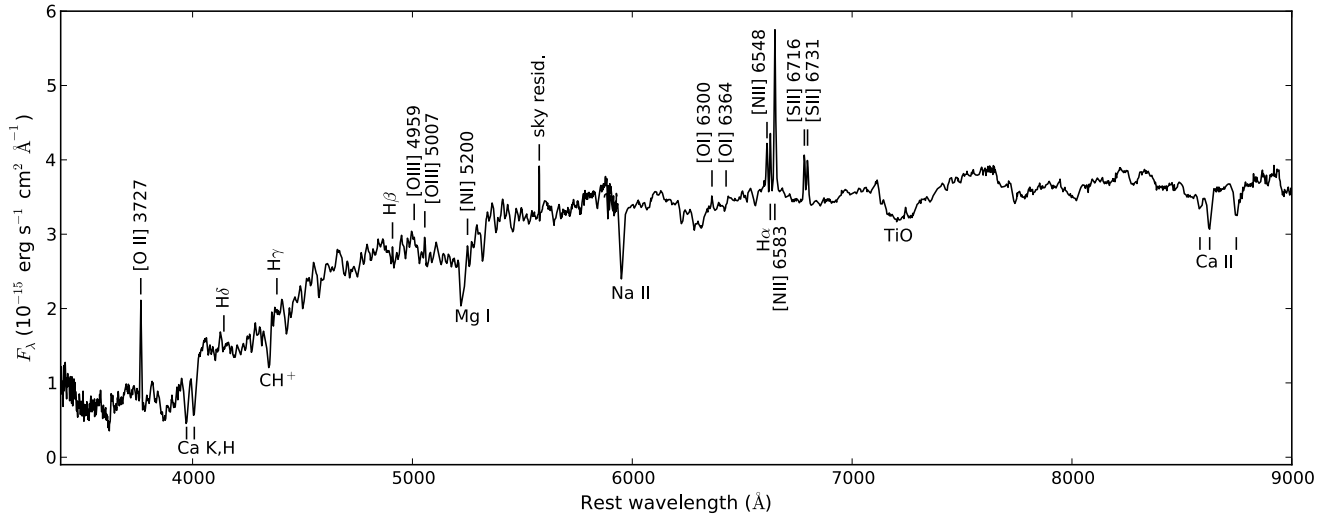}
\caption{Integrated spectrum of NGC~4696 from a 5\arcsec\,diameter circular aperture centered on the galaxy nucleus, plotted against the rest wavelength.}\label{fig:intspec}
\end{figure*}

Figure~\ref{fig:intspec} shows the integrated spectrum from a 5\arcsec-diameter circular aperture centered on the core of NGC~4696, prior to subtracting the galaxy continuum. Forbidden lines of \ion{N}{2}, \ion{O}{2} and \ion{S}{2}, and the H$\alpha$ recombination line are prominent, but the spectrum is dominated by a stellar continuum component that is typical of an early-type galaxy. The strength of the continuum increases toward longer wavelengths and exhibits absorption features characteristic of evolved stellar populations: e.g. Ca K and H lines (at 3934\,\AA~and 3969\,\AA, respectively), G band (CH$^{+}$; 4300\,\AA), blended Mg I b (5175\,\AA), Na I D blended doublet (5893\,\AA), and the broad TiO feature (7126\,\AA). In addition, the step in the average continuum intensity across the 4000\,\AA-break ($D_{4000}=1.9$) is typical of values in elliptical galaxies, including those of the low line-luminosity BCGs in the ROSAT `Brightest Cluster Sample' of \citet{Crawford:1999p1097}.

The galaxy spectrum shows no evidence for a blue stellar component that would indicate that significant star formation is occurring. This is in contrast to the blue and UV continuum excesses and large star formation rates estimated in high-luminosity BCGs: for example those in clusters Zw 3146 \citep[e.g.][]{Hicks:2005p6414}, Abell 1835 \citep{McNamara:2006p3331} and Abell 1068 \citep{Wise:2004p3017}.
It is consistent, though, with the general properties of the BGC population, which show reduced signs of star formation in less luminous BCGs \citep[e.g.][]{Allen:1992p1098,Crawford:1999p1097,Egami:2006p3830}, such as NGC~4696 itself.

We compare the radial surface brightness profile of the galaxy continuum with a Hernquist density model of the form 
\begin{equation}
\rho(r) = \frac{M_{t}}{2\pi}\frac{a}{r}\frac{1}{(a+r)^{3}},
\end{equation} 
following \citet{Lim:2008p9101}, where $M_{t}$ is the total galaxy mass and $a$ is a characteristic scale radius. We assume a constant $I$-band mass-to-light ratio of $M/L = 4.7$, taken from the correlation between stellar velocity dispersion and $I$-band $M/L$ presented by \citet{Cappellari:2006p8935}.

\begin{figure}
\epsscale{1.13}
\plotone{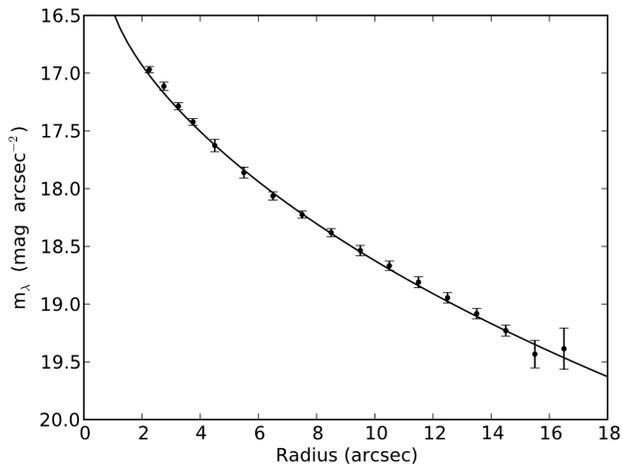}
\caption{Radial surface brightness profile of the stellar continuum emission in NGC~4696 at $\lambda=8000$\AA. Each point corresponds to the mean surface brightness of the pixels within a circular annulus. The error bars correspond to the dispersion in surface brightness of the pixels in each region.}\label{fig:galaxyprof}
\end{figure}

A simple minimum least-squares criteria is used to compare the model with the measured galaxy surface brightness at 8000\,\AA, averaged azimuthally between position angle 35\degr\,and 125\degr\,(i.e. in the quadrant of the WiFeS field where the dust absorption features are least prominent) and at radii from 2\arcsec\,to 17.5\arcsec. The best fit is obtained with a total mass $M_{t} = \scinot{(4.0\pm0.5)}{11}$\msol\,and scale radius $a=3.5$\,kpc. The measured and model radial surface brightness profiles are shown in Figure~\ref{fig:galaxyprof}.

\citet{Laine:2003p8500} fit the $I$-band surface brightness profile with a `Nuker profile' (a broken power law that has a turnover radius of 1.4\arcsec) in the central 10\arcsec, to fit the `core' seen in the profile data. We find that the Hernquist profile provides a better fit to the profile shape for our data beyond $r\sim10$\arcsec.

%%% referee (5)
Our estimated total stellar mass is different to the galaxy mass of \scinot{(11.6\pm0.1)}{11}\msol\,  presented by \citet{Rafferty:2006p2278} from $R$-band imaging. The procedure used by the authors to determine this mass is not described in sufficient detail to understand the nature of this discrepancy.

The stellar continuum is relatively redder closer to the galaxy core, as also observed by \citet{deJong:1990p3612}. This may be either the effect of dust extinction increasing towards the galactic center, or else the result of a greater mean metallicity in the stars.

%-------------------------------------------------------------------------
\subsection{Emission-line spectrum and line maps}\label{sec:linemaps}

\begin{figure*} 
\epsscale{1.05}
\plotone{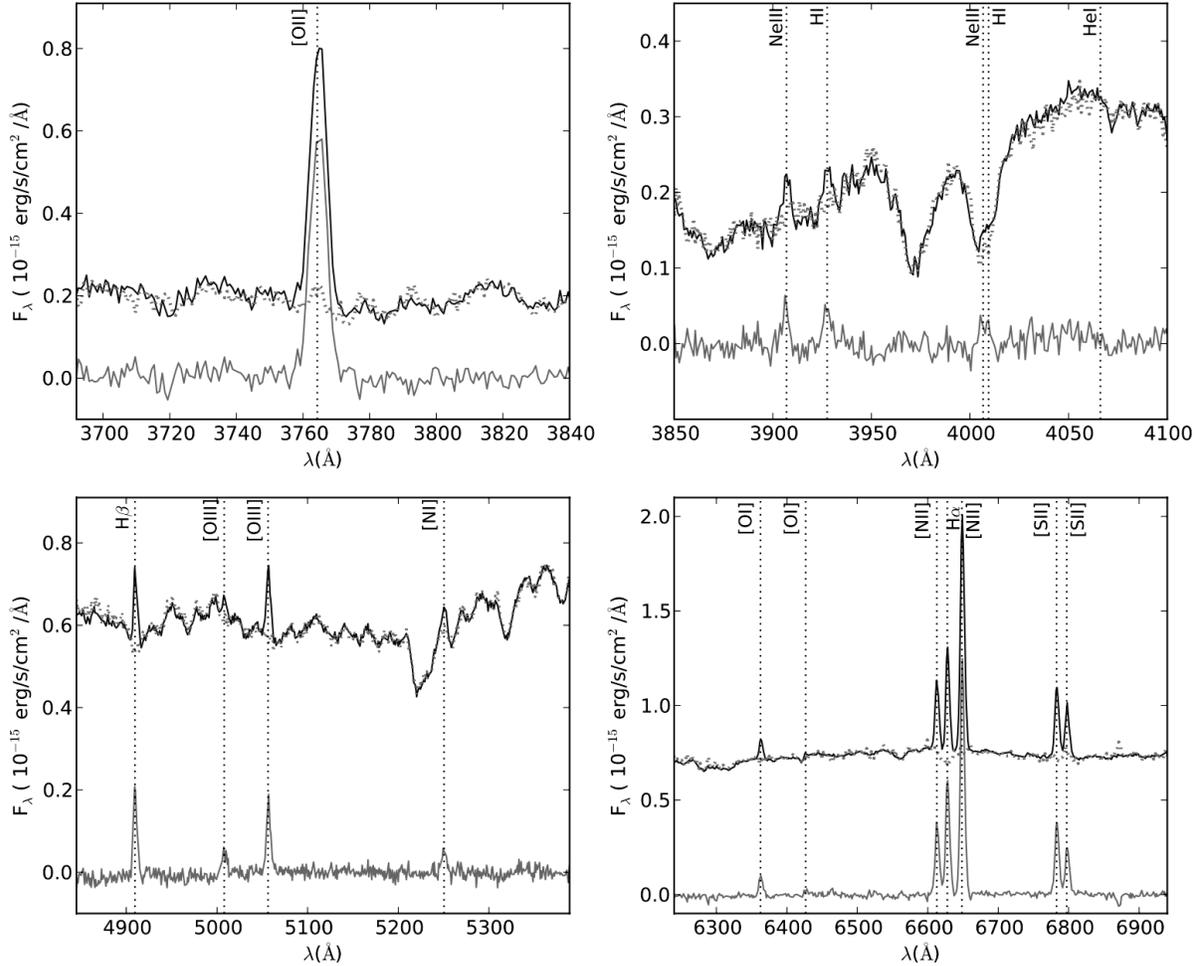}
\caption{These panels show regions of spectra near the strongest emission lines, extracted from a 3\arcsec-diameter aperture at the emission-line peak. The top (solid) line in each case is the raw integrated spectrum. Overplotted in dotted lines are the line-free galaxy spectrum used to subtract the stellar component. The lower traces show the (integrated) emission-line spectrum produced by subtracting the galaxy continuum.}\label{fig:contsub}
\end{figure*}

\begin{figure*}
\begin{center}\includegraphics[width=0.85\textwidth]{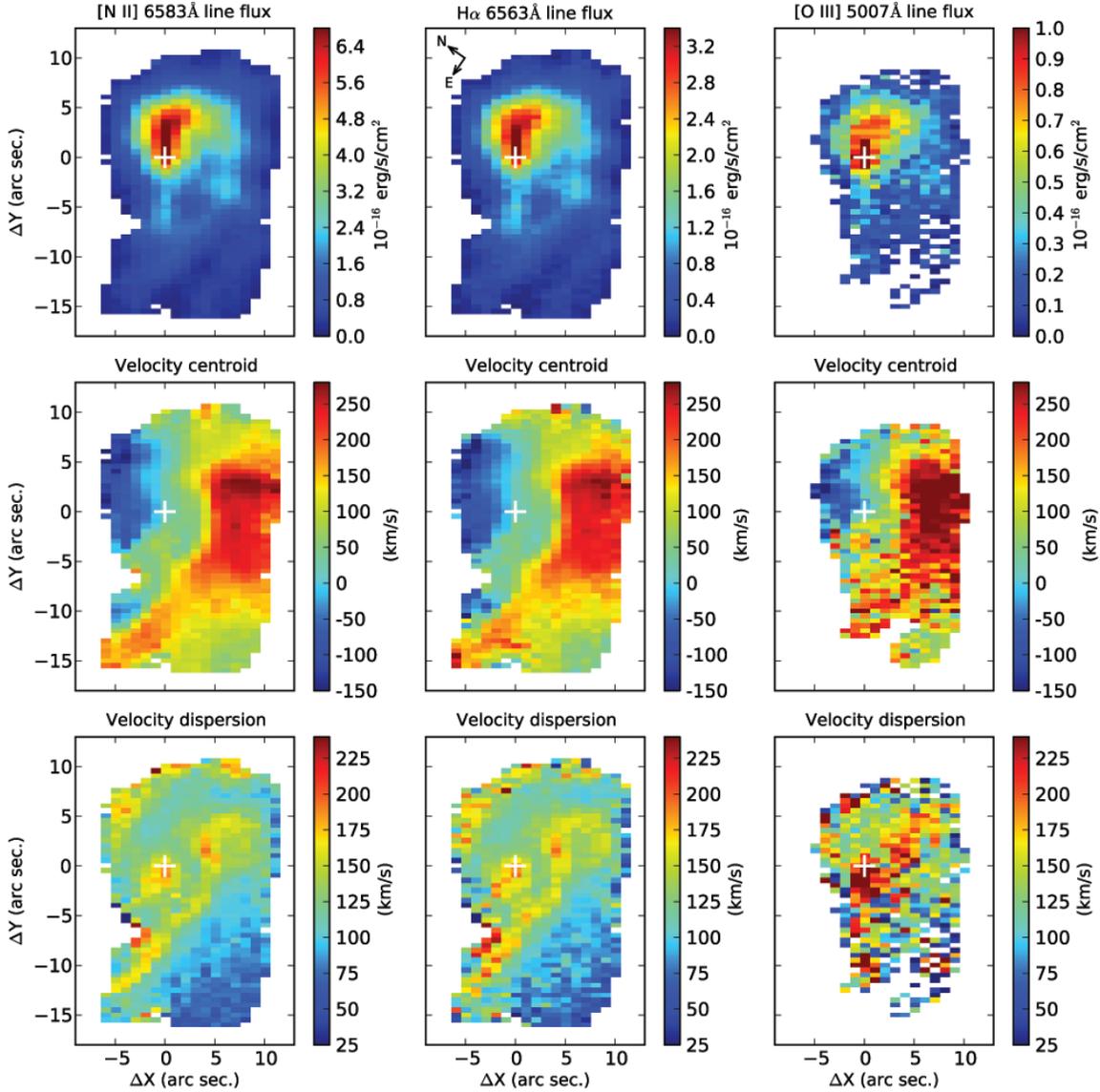}\end{center}
\caption{Measured pixel flux, velocity centroid, and velocity dispersion of the [\ion{N}{2}] $\lambda$6583\,\AA\,(left), H$\alpha$ $\lambda$6563\,\AA\,(center), and [\ion{O}{3}] $\lambda$5007\,\AA\,(right) lines. Note the strong increase in flux towards the center of the galaxy, the systematic rotation of the gaseous filaments, and the fairly uniform $\sim 120$\kms\, velocity dispersion throughout the emission-line system. The apparent increase in velocity dispersion in regions of strong velocity gradient is probably due to beam-smearing in these areas.}\label{fig:linemaps}
\end{figure*}

Figure~\ref{fig:contsub} presents emission-line spectra in the region of the filaments where the line emission is strongest, before and after continuum subtraction. The spectra are integrated over a 3\arcsec-diameter circular aperture centered approximately 3\arcsec\,north-west of the nucleus ($\Delta X=0.2$, $\Delta Y=3.0$\,\arcsec~in the reference frame of the images in Figure~\ref{fig:linemaps}). The reference galaxy spectrum is shown, scaled to match the data in each panel, together with the total spectrum and the continuum-subtracted result, illustrating the strongest emission-line detections. These spectra have not been corrected for reddening.

Table~\ref{tab:fluxes} lists integrated, continuum-subtracted and extinction-corrected line-flux measurements in the same 3\arcsec\, aperture, obtained from the Gaussian line profile fits. This table shows the emission features detected above the $3\sigma$ sensitivity level in the aperture. 
We correct the line flux measurements for extinction using the reddening law of \citet{Osterbrock:1989p9491}: the relative extinction at the wavelength of H$\alpha$ is $A_{\rm H\alpha}/A_V =  0.744$ and at the wavelength of H$\beta$ is $A_{\rm H\beta}/A_V = 1.14$. In addition, we adopt a value of $R = A_V / E(B-V) = 3.1$.
To determine the total extinction, we assume that the intrinsic emitted flux ratio $F$(H$\alpha$)/$F$(H$\beta)$ = 3.1. This is somewhat higher than the classical Case B recombination value, but is appropriate for active galaxies, including LINERs \citep{Veilleux:1987p889}, in which there is a collisional excitation component in the H$\alpha$ flux.
From the dust maps of \citet{Schlegel:1998p10426}, the reddening contribution from our Galaxy in the direction of NGC~4696 is $A_V\sim0.34$.

\begin{table}[b]
\begin{center}
\caption{Extinction-corrected flux ratios for emission lines integrated over a 3\arcsec-diameter aperture at the emission peak, relative to the measured H$\alpha$ line flux: $F$(H$\alpha$) = \scinot{(6.7\pm0.7)}{-15}\flux. \label{tab:fluxes}}
\begin{tabular}{p{3.5cm}p{2.2cm}}
\tableline \tableline
Emission line			&	$F(\lambda)/F(\rm H\alpha)$  \\
\tableline
$[$O II$]$ 3726+3729	&	$	1.49	\pm	0.06	$		\\
H$\gamma$ 4340		&	$	0.13	\pm	0.01	$		\\
H$\beta$ 4861			&	$	0.32	\pm	0.01	$		\\
$[$O III$]$ 4959		&	$	0.06	\pm	0.01	$		\\
$[$O III$]$ 5007		&	$	0.31	\pm	0.01	$		\\
$[$N I$]$ 5198+5200	&	$	0.12	\pm	0.01	$		\\
$[$O I$]$ 6300			&	$	0.20	\pm	0.01	$		\\
$[$N II$]$ 6548			&	$	0.65	\pm	0.01	$		\\
$[$N II$]$ 6583			&	$	2.05	\pm	0.02	$		\\
$[$S II$]$ 6716			&	$	0.61	\pm	0.01	$		\\
$[$S II$]$ 6731			&	$	0.41	\pm	0.01	$		\\
\tableline
\end{tabular}
\end{center}
\end{table}

%%% referee (5)
The resulting extinction measured in the 3\arcsec-diameter aperture at the line peak is $A_{V} = 0.74\pm0.07$\,mag. The error is obtained from the uncertainty in the measured H$\alpha$/H$\beta$ flux ratio integrated over the aperture. The measurements represented in Table~\ref{tab:fluxes} have been corrected for extinction of this magnitude. 
The measured H$\alpha$ line flux in the aperture is \scinot{(4.0\pm0.4)}{-15}\flux, and the extinction-corrected result is \scinot{(6.7\pm0.9)}{-15}\flux. The error here is determined from the uncertainty in the flux calibration and the extinction. 
%%% referee (6)...
Line fluxes for the other detected emission lines are presented as ratios to the H$\alpha$ flux in Table~\ref{tab:fluxes}. Estimated uncertainties in the flux ratios, derived as described in Section~\ref{sec:linefits}, are also shown in the table.

The top-right panel of Figure~\ref{fig:contsub} indicates possible detections of the [\ion{Ne}{3}] $\lambda$3869\AA, H$\zeta\, \lambda$3889\AA\, and blended [\ion{Ne}{3}] $\lambda$3967\AA\,+ H$\epsilon\, \lambda$3970\AA\, lines, which are close to or below the 3$\sigma$ detection threshold and could not be fit in individual pixels.

Figure~\ref{fig:linemaps} presents maps of the emission-line properties derived from the Gaussian profile fits for three strong emission lines in the spectra: the line flux in the top panel (no reddening corrections have been applied), the velocity of the line centroid in the middle panel, and the derived velocity dispersion in the bottom panel. The cross marks the position of the galaxy nucleus, at the peak of the continuum emission surface brightness distribution.

%---------------------------------------------------------------------------------------
\subsection{Emission-line region morphology}\label{sec:morph}

\begin{figure*}
\epsscale{0.95}
\plotone{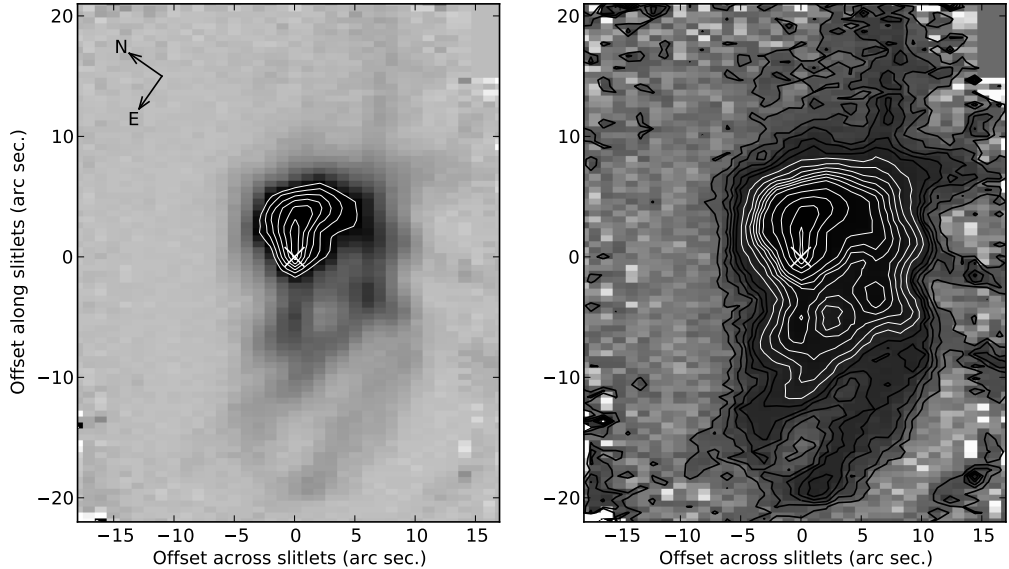}
\caption{Maps of integrated \ion{N}{2} 6548, 6583\AA\,+ H$\alpha$ line emission from the continuum-subtracted NGC~4696 data cube. The image on the left shows the detail of the inner structures with a linear grayscale and contours of integrated flux. The right panel shows the same image with histogram-equalization of the levels to highlight the faint outer filaments.}\label{fig:integrated}
\end{figure*}

The distribution of the detected line emission from the filaments of NGC~4696 is illustrated in Figure~\ref{fig:integrated}, which shows a projection along the wavelength axis of the data cube over the [\ion{N}{2}] $\lambda\lambda$6548,6583\AA\,and H$\alpha$ lines. 
The structures detected here can be matched to the features seen in narrow-band imaging by \citet{Crawford:2005p334}. The detected filaments extend to the east, south, and west of the galaxy center. Some filaments have a generally radial orientation with respect to the galaxy, but there are curved arcs, bars, and filaments in the structures that are extended at orientations that are much more nearly tangential. 

The most prominent feature is the bright filament that spirals from the nucleus to the north-west, and in the outer regions curves around the galaxy core to the east. This is the structure that was identified in early imaging of the gas and dust by \citet{Fabian:1982p2130}, \citet{Jorgensen:1983p3750}, \citet{NrgaardNielsen:1984p9791}, and \citet{Sparks:1989p2171}. The surface brightness of the filament decreases with distance from the galaxy nucleus along its length (other than at a brighter knot that appears at the intersection with another filament). 
The outer length of this main filament, to the south of the galaxy core, coincides closely with the prominent dust lane that is seen clearly in the HST \citep{Laine:2003p8500} and $B-I$ color \citep{Crawford:2005p334} images of the galaxy. In the HST images, this lane of dust appears to be part of a continuous structure that spirals around the galaxy center at small radii to the north-east and with increasing radius to the west and south, passing through the brightest regions of the line emission.
Lanes of absorption also coincide with outer filaments to the south and west of the main filament.
The association of the line emission and dust absorption features indicates that the observed system of gas and dust is inclined towards us, is on the near side of the galaxy, and has an intrinsically one-sided  spiral distribution about the galaxy. 

\begin{figure}[b]
\epsscale{0.85}
\plotone{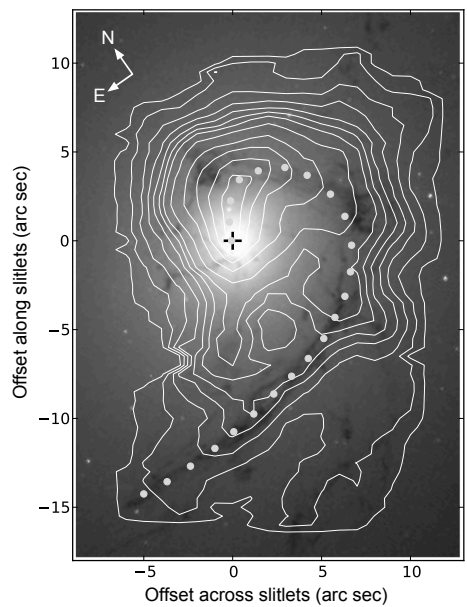}
\caption{An HST image of the inner regions and dust lane in NGC~4696, obtained from the Hubble Legacy Archive (HST proposal 9427; dataset: HST\_9427\_06\_ACS\_WFC\_F435W), with contours of optical emission ([\ion{N}{2}] flux from our data) overlaid. The main optical filament is marked by the white points, which were selected by eye to follow the ridge of emission along the main filament. This ridge line, initially along the outer edge of the dust lane, progressively moves inside the dust spiral towards the nucleus.}\label{fig:dustlane}
\end{figure}

The peak in surface brightness of the extended emission occurs approximately 3\arcsec\,(0.6\,kpc) to the north-west of the galaxy center. This peak occurs inside the curve of the dust lane. The spatial association of the main filament with the dust lane is very clear in the outer extent of the filaments. Close to the nucleus, however, the main dust spiral departs from the ridge line of the line emission along the filament. As seen in Figure~\ref{fig:dustlane}, the filament of optical emission bends toward the nucleus from the north-west, whereas the dust spiral continues around to the north of the core and finally turns towards the nucleus from the north-east. If the ridge of emission tracks a shock front, as proposed below, this might indicate that the shock front is progressively propagating through the dust lane as the dusty gas spirals in towards the nucleus.

The morphology of the emission from weaker lines such as [\ion{O}{1}] and [\ion{N}{1}] is consistent with that of the strong emission lines (hence are not pictured in Figure~\ref{fig:linemaps}). To the degree that our spatial resolution reveals, it appears that all the optical forbidden and hydrogen recombination emission lines originate in the same gas. 

%-------------------------------------------------------------------------
\subsection{Line luminosity and mass of filaments}\label{sec:linelum}

The integral-field data permit us to determine the total emission-line luminosity of the filaments, which is useful to constrain both the mass and energy budgets. 
The flux measurements from the emission-line fits are summed over all the pixels where the value is greater than the 3$\sigma$ detection limit to obtain a total line flux. 

%%% referee (5 ii):  
The total flux detected in H$\alpha$ is \scinot{(5.7\pm 0.6)}{-14}\flux\, prior to correcting for extinction. 
The reddening in the observations is estimated from measured H$\alpha$/H$\beta$ line flux ratios as described in Section~\ref{sec:linemaps}. Since H$\beta$ is relatively faint, we do not detect it over the same spatial extent as we measure the total H$\alpha$ flux. This restricts our ability to map the spatial variation of the extinction.
In spatial pixels where both lines are detected above the 5$\sigma$ detection limit, the extinction correction is determined from the H$\alpha$/H$\beta$ line ratio measured in that pixel. Everywhere else, we assume an average extinction level of $A_V=0.4$\,mag.
This value is determined from the integrated line fluxes over the region where both H$\alpha$ and H$\beta$ are detected. The line luminosity is then derived by summing the individually extinction-corrected fluxes from each pixel.
 
This results in an extinction-corrected, integrated H$\alpha$ flux of \scinot{(7.9\pm 0.9)}{-14}\flux. At a distance of 44\,Mpc, the H$\alpha$ line luminosity is then \scinot{(1.8\pm0.2)}{40}\lum. 
This is consistent with the lower-limit estimate of \citet{Crawford:2005p334} from narrow-band imaging (saturation in their images allowed only a lower limit to be obtained): $L({\rm H\alpha})>\scinot{1.5}{40}$\lum. The total [\ion{N}{2}]\,$\lambda$6583\,\AA\, luminosity from our data is \scinot{(3.7\pm0.4)}{40}\lum, similar to the value of $\sim$\scinot{5}{40}\lum\, measured by \citet{Fabian:1982p2130} for the [\ion{N}{2}]\,$\lambda$6583\,\AA\, flux. 
In earlier narrow-band work, \citet{Sparks:1989p2171} estimated an approximate combined H$\alpha$ + [\ion{N}{2}]\,$\lambda$6548+6583\,\AA\, luminosity of \scinot{9}{40}\lum, and given the inherent difficulties in calibrating such measurements and accurately extracting only the line emission, this is in reasonable agreement with our result.

An estimate of the mass of emitting ionized hydrogen can be made using the H$\alpha$ emission-line luminosity, $L({\rm H}\alpha)$ \citep{Osterbrock:1989p9491}:
\begin{equation}
M_{\rm H^{+}} = \frac{L({\rm H}\alpha)\,m_{p}}{n_{e}\,\alpha_{\rm{H}\alpha}^{\rm eff}\,h\,\nu_{{\rm H}\alpha}} \label{eq:massH}
\end{equation}
where we adopt the effective recombination coefficient for H$\alpha$ line emission $\alpha_{\rm{H}\alpha}^{\rm eff} = \scinot{1.17}{-13}$\,cm$^{3}$\,s$^{-1}$, from \citet{Osterbrock:1989p9491} for Case B emission at $T=10^{4}$\,K. Then
\begin{equation}
M_{\rm H^{+}} \approx \scinot{4.3}{5}\left( \frac{100\,{\rm cm}^{-3}}{n_{e}}\right)\,\hbox{\msol}. 
\end{equation}
Thus, assuming an average electron density of $\sim100$\,cm$^{-3}$, we find a total mass of ionized hydrogen in the filaments of $M_{\rm{H}^{+}} \sim \scinot{(4.3\pm0.5)}{5}$\,\msol.

The mass and density  can be used to define a volume filling factor $\varepsilon$ for the ionized gas in the filaments. The volume filling factor for material in a structure is defined as the fraction of the encompassing volume that is filled by material at the estimated density. An assumption about the geometry of the emitting region is required. 
We can calculate a filling factor for the ionized material within a spherical volume that encompasses the filaments: 
\begin{equation}
\varepsilon = \frac{M_{H^{+}} / \rho } {\case{4}{3}\,\pi R^{3} }
\end{equation}
where $M_{H^{+}}$ is the mass of ionized gas derived from the H$\alpha$ flux, $\rho$ is the mass density of the emitting gas, and $R$ is the radius of a sphere encompassing the emitting region.
Assuming an electron density $n_{e}\sim 100$\,cm$^{-3}$, $R \sim 2.9$\,kpc and the ionized gas mass determined above, the volume filling factor is $\varepsilon \approx 10^{-6}$.
This is comparable to values found for other filament systems \citep[e.g.][]{Ogrean:2010p10469,Hatch:2007p2057}, and is consistent with the H$\alpha$-emitting gas being only a small fraction of the emitting material, located in a thin layer on the surface or within the volume of a larger mass of gas. 

This larger volume is unlikely to be spherical, so a more appropriate estimate might be obtained  within a more constrained region of the filaments. We assume the emitting material from the 3\arcsec-diameter aperture used in Section~\ref{sec:linemaps} and Table~\ref{tab:fluxes}, at the line peak, is distributed over a spherical volume of radius $R = 1.5\arcsec \sim \scinot{9.7}{20}$\,cm. From Equation~\ref{eq:massH}, the estimated emitting mass of ionized gas in this aperture is $M_{H^{+}} = \scinot{(3.6\pm0.5)}{4}$ and the volume filling factor is $\varepsilon \approx 10^{-4}$, with $n_{e}\sim100$\,cm$^{-3}$. This is still small, indicating a clumpy or filamentary distribution, but may more accurately reflect the volume fraction of the main filament that is occupied by ionized gas.

The total gas mass is an important parameter, but is difficult to accurately ascertain. We discuss below various estimates from the literature of the masses of the gas and dust in the filaments. These provide an indication of the quantities of material present.
\citet{Sparks:1989p2171}, for example, studied the dust absorption in optical imaging of the galaxy and calculate a mass of dust in the main dust lane $>\scinot{2}{7}$\msol. This would then imply, using a Galactic gas to dust mass ratio, that the atomic component of the filaments has $M\sim\scinot{2}{9}$\msol.  

Using the color gradient in the stellar continuum, \citet{deJong:1990p3612}  estimate the mass of dust present in the galaxy and dust lane from IRAS 100\,$\mu$m observations, by assuming that the gradient is entirely due to reddening rather than metallicity changes, and calculating the mass of dust required to produce it. They obtain a total dust mass of $\sim\scinot{0.2-5}{6}$\msol, with approximately 10\% of this mass located in the dust lane and the rest distributed over the central $\sim$10\,kpc of the galaxy. Assuming a Galactic gas-to-dust ratio, this implies a total gas mass on the order of $M\sim\scinot{3}{8}$\msol. 

Such estimates of dust mass cannot fully take into account the total gas mass, if the dust clouds become optically thick and therefore cool. The mass in BCG line-emitting filaments is believed to be dominated by molecular material in such cloud cores. \citet{Edge:2001p4022} have measured CO emission in a sample of luminous cool-core galaxies that is consistent with the presence of $10^{9}-10^{11.5}$\msol\, of molecular gas at $T\sim40$\,K in these galaxies. Using Figure 10 of their paper, along with the inferred mass deposition rate for NGC~4696 of 20\msol\,yr$^{-1}$ from \citet{Fabian:1982p2130} then implies a molecular gas mass of $M\sim\scinot{2}{9}$\msol\, in NGC~4696, in close agreement with the value inferred by \citet{Sparks:1989p2171} from the dust obscuration.
In NGC~4696, \citet{Johnstone:2007p2172} derive a mass of warm H$_{2}$ of \scinot{1.3}{5}\msol, from emission within their Spitzer spectrograph aperture and estimate that approximately 10 times more mass may exist at cooler temperatures. 

As previously, we adopt an average reddening of $A_V\sim0.4$ along the extent of the main filament. Assuming a Galactic gas-to-dust ratio of $N_{H}/E(B-V)\sim\scinot{5.8}{21}$\,cm$^{-2}$\,mag$^{-1}$ \citep{Bohlin:1978p10470}, a length of 32\arcsec, and width of 2\arcsec\,over which the emission in the filament is measured, we can make a rough estimate of the mass of material obscuring the line-emission region of $M \sim 10^{7}$\msol. This estimate is a lower limit for total mass because the ionized gas cannot be located entirely behind a dusty foreground screen, but is more likely mixed in with it. Over the total area of the line-emitting filaments, using the same average reddening, the resulting limit on the total mass is $M>\scinot{8}{7}$\msol.

Although these mass estimates are approximate and vary over a large range, they clearly indicate that there is a large mass of material in the extended filaments, of order a few times $10^{9}$\msol. It is most likely that this mass is largely in a cool, dusty, and dense molecular phase.

%--------------------------------------------------------------------------------------
\subsection{Kinematics of the emitting gas} \label{sec:kinematics}

There have been a number of long-slit spectroscopic studies of the kinematics of the gas in NGC~4696. Studies by \citet{Sparks:1997p2056} and \citet{deJong:1990p3612} revealed smoothly varying line-of-sight velocities with values changing by $\sim$400\kms~in the main filament.  In addition, \citet{Sparks:1989p2171} detect sodium absorption features in continuum-subtracted long-slit spectra that have similar kinematics to the emission-line gas along both position angles observed. The measured line-of-sight velocities are between 0 and 450\kms. There is a cold, neutral component that is associated with the ionized filaments. They comment that this would not be expected in a cooling-flow model, but is consistent with a merger. 

Our integral-field spectroscopy greatly improves our ability to map the velocity dispersions and kinematics of the line-emitting gas. The middle panels of Figure~\ref{fig:linemaps} show the line-of-sight velocity of the emitting material measured from the Doppler shift of the Gaussian line profile centroids, relative to the galaxy. The bottom panel shows the line-of-sight velocity dispersion from the fitted profile widths, after correcting for the instrumental width ($\sigma=41$\kms).

The velocity dispersion detected across the main filament is relatively uniform, with most of the material having velocity dispersions of $\sim120\pm10$\kms\, (see Figure~\ref{fig:linemaps}).  
Ridges of higher dispersion (160-200\kms) appear in the maps, but the dispersions in these regions are consistent with  `beam-smearing'  caused by our finite spatial resolution, which results in an increase in apparent velocity dispersion in regions of a steep spatial gradient in line-of-sight velocity. The measured velocity dispersions from the H$\alpha$ and [\ion{N}{2}] lines are effectively identical.

The velocity dispersions in the main filament are significantly lower than the rotational/bulk velocities of the gas: without correcting for inclination, the ratio of rotation to turbulent velocity ranges from 1 to 2 in this structure. The central stellar velocity dispersion in the galaxy is also a factor of two larger than the line-of-sight velocity dispersion of the line emission. 
The velocity dispersion is much larger than the thermal broadening ($\sim$7\kms\,at 10$^{4}$\,K for hydrogen) and is therefore either generated by shocks or by the superposition of large-scale random turbulent motions.

The main filament has broader lines and higher line-of-sight velocities than the outer filament arm to the south, in which the average velocity dispersion ranges from $\sim$60\kms\,to $\sim$90\kms\,and the line-of-sight velocity from $\sim$100\kms\,to $\sim$150\kms. This outer filament also has lower surface brightness. 

The velocity distribution of the line-emitting gas shows an ordered structure with a component of rotation in the nebula. The velocity maps derived from the line centroid positions for other ions are essentially identical to the [\ion{N}{2}] and H$\alpha$ velocity maps shown in Figure~\ref{fig:linemaps}, indicating that the hydrogen recombination and forbidden lines in the optical spectrum originate from the same gas.

The emission from the ionized gas is largely redshifted to the south of the galaxy center and blue-shifted to the north. There is a smooth variation in the line-of-sight velocity along the principal filament that traces the dust spiral. The velocity increases inward along the filament as it bends around the galaxy to the south of the core, and then decreases to become blue-shifted as the filament curves to the north past the galaxy nucleus. As a result, the kinematics of the gas in this filament resemble orbital motions: if the material is infalling, then the filament is inclined so that the outermost end is nearest to us and the material recedes along the filament to the south of the core, before turning around on the far side of the galaxy core onto an approaching trajectory to the north of the nucleus. 

At the outer end of the main filament at a radius of 13\arcsec\, (2.7\,kpc), the average line-of-sight velocity of the emitting gas is 180\kms. This velocity increases to 280\kms\, at a radius of 6.5\arcsec\,(1.4\,kpc) in the filament to the south-west. The maximum blueshifted velocity is 150\kms\, at a radius of 7\arcsec\,(1.5\,kpc) to the north of the core. There is a total velocity range of 430\kms\, in the measured line centroids, similar to the value measured by \citet{Sparks:1997p2056}.

Assuming a Keplerian orbit with radius $r=1.5$\,kpc and circular velocity $v=300$\kms, the enclosed mass would be $\scinot{3.1}{10}$\msol. The mass profile derived from the continuum surface brightness profile of the galaxy in Section~\ref{sec:galcont} indicates that there is a stellar mass of \scinot{3.6}{10}\msol\,inside this radius, in excellent agreement with our kinematic estimate, and showing that the contribution of dark matter to the total mass in this central region must be fairly small. The orbital timescale implied by material at a galactic radius of 1.5\,kpc travelling at velocity $\sim300$\kms\, is $\sim\scinot{3}{7}$\,yr. 
Taking the total gas mass (from Section~\ref{sec:linelum}) as $2\times 10^{9}$\msol, the kinetic energy of the gas  would be $1/2M v^{2} \sim \scinot{1}{57}$\,erg. If this energy is radiated away by shocks on the timescale of a single orbit, the luminosity produced would be \scinot{2}{42}\lum. The H$\alpha$ line luminosity from our line measurements is \scinot{(1.8\pm0.2)}{40}\lum, and from the shock models presented below, this is about 0.7\% of the total luminosity. Thus, our estimated total shock luminosity is  $\sim\scinot{2.6}{42}$\lum, in good agreement with our kinematic estimate of this quantity.

 %----------------------------------------------------------------------------
\subsection{Spatial variation of excitation} \label{sec:excitation}

\begin{figure*}
\epsscale{0.84}
\plotone{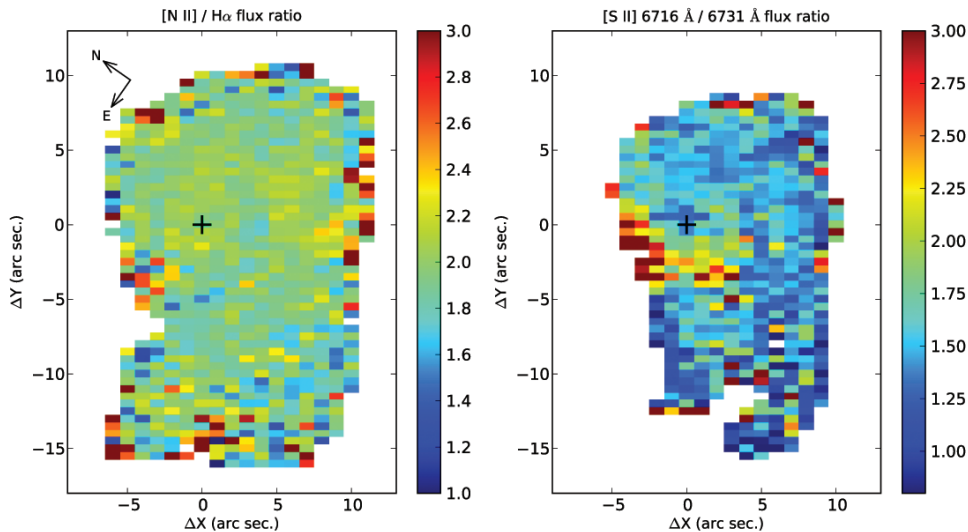}
\caption{Flux-ratio maps of the [\ion{N}{2}] to the H$\alpha$ recombination line emission (left), and for the density-sensitive [\ion{S}{2}] doublet (right).}\label{fig:fluxratios}
\end{figure*}

Previous long-slit spectroscopy of NGC~4696 has shown that the extended filaments produce emission with a high ratio of [\ion{N}{2}] to H$\alpha$ flux \citep{Lewis:2003p2494}. We confirm this result. Figure~\ref{fig:fluxratios} shows that the ratio of [\ion{N}{2}]/H$\alpha$ is roughly constant at $2.0\pm0.1$ with remarkably little variation in the value ($\sim$5\%) across the inner filaments of the galaxy.
This average value of the [\ion{N}{2}]/H$\alpha$ flux ratio and the H$\alpha$ luminosity of the system reported in Section~\ref{sec:linelum} are consistent with other BCG emission systems of similar luminosity \citep{Crawford:1999p1097}. These results show that the characteristic line ratios such as the high [\ion{N}{2}] flux relative to H$\alpha$ and dominance of the other low-ionization species, are not simply nuclear phenomena confined to the galactic core, but persist over the extent of the detected emission in the filaments.  In addition, the uniformity of the emission-line flux ratios across the inner filaments implies that whatever mechanism excites this spectrum is acting over the full extent (25\arcsec\,or 5\,kpc) of the detected filaments. 

This finding contrasts with results from other BCG filament systems. For example, \citet{Hatch:2006p1101} placed a spectrograph slit along several of the filaments in NGC~1275 and found the [\ion{N}{2}]/H$\alpha$ ratio decreases along the filaments with distance from the galaxy. 
The authors suggest that this may be the result of either decreasing metallicity or changing excitation, such as greater specific star formation activity closer to the nucleus. Unlike NGC~4696, NGC~1275 is a very line-luminous system among BCGs and there is clear evidence of star-forming clouds associated with the extended filaments.

 %%% referee(12)
\citet{Edwards:2009p11292} compare optical IFU observations of the extended emission in a sample of cooling-flow and non-cooling-flow BCGs. They find diverse emission properties and morphologies among their sample, including a range of behavior in the variation of the [\ion{N}{2}]/H$\alpha$ line flux ratios. In general, they attribute lower [\ion{N}{2}]/H$\alpha$ ratios to regions where young stars contribute most of the ionising flux. They conclude that the ionization mechanisms that produce the optical line emission are not uniform amongst the sample and represent a complex interplay between an AGN, cooling cluster gas, and star formation that may be triggered by AGN outbursts and/or tidal interactions with other galaxies. 

In IFU observations of a sample of other BCGs with a range of H$\alpha$ luminosities, \citet{Hatch:2007p2057} find different degrees of [\ion{N}{2}]/H$\alpha$ variation in the observed systems. For example, in RXJ 0821+0752 there is a distinct correlation between the low [\ion{N}{2}]/H$\alpha$ and H$\alpha$ flux, whereas in the BCG of Abell 2390 there is less variation, but in general higher [\ion{N}{2}]/H$\alpha$ ratios are associated with dust lanes and lower ratios with knots of blue emission. This may also reflect the trend amongst BCGs to display weaker forbidden emission in high-luminosity systems with abundant star formation (as evidenced by blue knots). 
 \citet{Wilman:2006p3338} found little variation in the flux ratios over the measured extent of nebulae surrounding four high-luminosity systems. If, as suspected, such systems are dominated by more widespread regions of star formation, the emission-line ratios may be simply consistent with photoionization by hot stars. On this hypothesis, they suggested that in lower-luminosity systems the flux ratios might show more spatial variation with higher [\ion{N}{2}]/H$\alpha$ ratios away from the less-abundant star-forming clumps. 
 
NGC~4696 may represent the low-luminosity extreme of this model, as there are no detectable star-forming regions associated with the filaments. In this case, the underlying low-ionization excitation mechanism dominates everywhere. M87, low line-luminosity BCG of the Virgo Cluster, may be a similar example, though in this case observations resolve the core of the galaxy and the emission-line ratios associated with the extended emission differ from those of the nucleus. The extended filaments produce a typically LINER-type spectrum \citep{Ford:1979p5098}, found by \citet{Dopita:1997p2854} to be consistent with shock models, whereas the higher-ionization emission from the core is consistent with excitation by photoionization \citep{Sabra:2003p2665}.

The [\ion{O}{3}]/H$\beta$ ratio is a powerful diagnostic of the excitation of the gas. This ratio is also measurable in the central regions of the filaments in our data, though the signal-to-noise ratio is much reduced as a result of the overall weaker line emission. We find little evidence of a variation in the value over the inner filament. However, there is evidence for a slight increase in the ratio at the galaxy core, where the value may be enhanced by radiation from the (weak) AGN in this system. 
%%% referee (8)
This effect can be seen in the line flux maps of Figure~\ref{fig:linemaps}, as the peak of the [\ion{O}{3}] emission is closer to the galaxy core than the peak of the hydrogen lines.
%%% referee (11)
However, over most of the extent of the filaments, the line ratios, including [\ion{O}{3}]/H$\beta$, do not indicate a dominant contribution from AGN photoionization. The main filament uniformly produces a low-ionization (LINER-like) emission spectrum (see Section~\ref{sec:diagd}.
We do not detect \ion{He}{1}\,$\lambda$5876\AA, which is sometimes seen in more H$\alpha$-luminous BCG systems \citep[e.g.][]{Sabra:2000p2757}.

The intrinsic velocity dispersion in the ionized gas causes the [\ion{O}{2}]\,$\lambda3729,26$\,\AA\,doublet to be unresolved in our spectra. However, our wavelength range includes another density-sensitive pair of collisionally-excited lines: [\ion{S}{2}]\,$\lambda$6716,31\,\AA. Figure~\ref{fig:fluxratios} maps the ratio of the flux in the 6731\,\AA\,component to that in the 6716\,\AA\,line.
The average value of the [\ion{S}{2}] doublet flux ratio over the central filament is $1.5\pm0.1$, which is consistent with densities close to or within the low-density limit ($\lesssim300$\,cm$^{-3}$). Again the [\ion{S}{2}] flux ratio is uniform over the brightest regions in the filaments, with a 1$\sigma$ variation of 0.2 in the value.

%-------------------------------------------------------------
\subsection{Diagnostic diagrams}\label{sec:diagd}
The most commonly-used optical diagnostic line-flux diagrams were developed by \citet{Baldwin:1981p3962} and \citet{Veilleux:1987p889}. These employ strong lines that are close in wavelength to decrease the effect of reddening on the ratios. Such diagnostic diagrams are often used in classifying galaxies and identifying AGN versus starburst-dominated systems by distinguishing different ionization mechanisms \citep[e.g.][]{Kewley:2006p1511}.

\begin{figure}
\epsscale{1.03}
\plotone{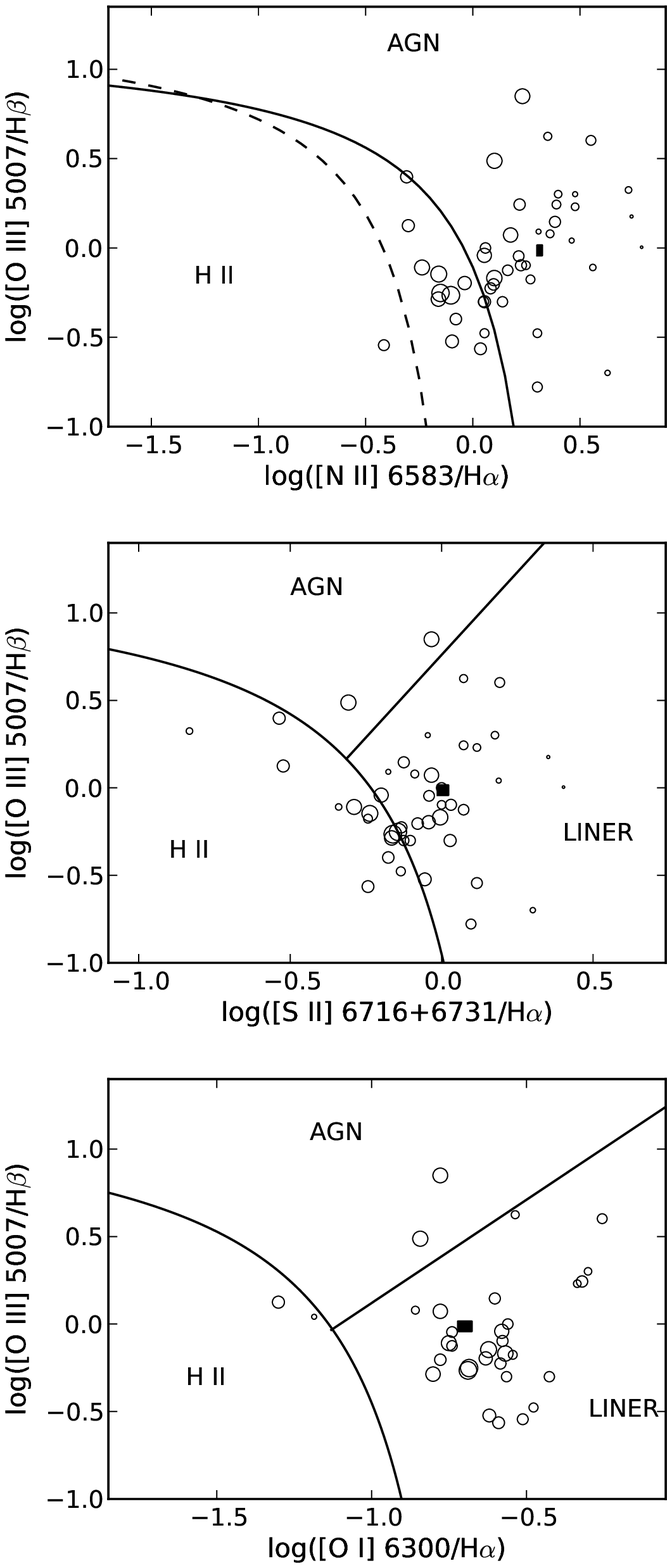}
\caption{The integrated line flux ratios from the 3\arcsec-diameter aperture located at the line peak in the main filament of NGC~4696 are plotted here on optical diagnostic diagrams as filled rectangular points, with side lengths that represent the uncertainties. For comparison, the flux ratios for galaxies in ROSAT BCS clusters presented in \citet{Crawford:1999p1097} are also plotted. For the BCS points, the symbol size corresponds to the H$\alpha$ luminosity of the system.}\label{fig:BPT}
\end{figure}

In Figure~\ref{fig:BPT}, we plot the flux ratios from Table~\ref{tab:fluxes} at the emission-line peak on the common diagnostic diagrams.  This figure shows the regions occupied by AGN, \ion{H}{2} regions and LINERs, as defined by the classification scheme of \citet{Kewley:2006p1511}, developed from their earlier work \citep{Kewley:2001p2767}. The typical line ratios from Table~\ref{tab:fluxes} for NGC~4696 are plotted along with the integrated line ratios measured by \citet{Crawford:1999p1097} for a sample of X-ray-selected BCGs. The relative sizes of the points represent the relative H$\alpha$ luminosities measured in that study. These demonstrate clearly the LINER-like properties of the filament spectrum of NGC~4696, typical of a low H$\alpha$-luminosity system. Thus any conclusions that apply to the mode of excitation of NGC~4696 may be applicable to other low-luminosity BCGs.

%----------------------------------------------------------------------------------------------------------------------------
\section{NGC~4696 as a minor merger}\label{sec:merger}  

There is extensive work in the literature that addresses the question of how the mass of brightest cluster galaxies is assembled, including the fraction of the mass that may be obtained through merger and accretion events. The bulk of the mass of brightest cluster ellipticals is believed to have been assembled in this way. BCGs are located in the densest galactic environments and the galaxy luminosities are weakly correlated with the cluster richness, X-ray luminosity, and velocity dispersion \citep[e.g.][and references therein]{Schombert:1987p10771}, signs that they are products of their host cluster environment. There is also a high frequency of multiple nuclei amongst BCGs \citep[e.g.][]{Hoessel:1980p10770}.

We have already accumulated a variety of evidence to support the idea proposed by \citet{Sparks:1989p2171} that the filaments of NGC~4696 represent the infalling remnants of a minor merger with a neighboring gas-rich cluster galaxy:
\begin{itemize}
\item{The kinematics of the emission-line gas show that the main spiral-shaped filament in NCC~4696 is a coherent structure consisting of material with infalling, orbital motions.}
\item{The one-sided distribution of the emission-line streams is also readily explained by the accretion of another galaxy, infalling on a timescale comparable to the orbital timescale.}
\item{As \citet{Sparks:1989p2171} point out, the presence of large quantities of dust, and molecular, neutral, and ionized gas in the filaments is also consistent with a merger origin.}
\item{As discussed in Section~\ref{sec:linelum}, there is evidence that the mass of material associated with the filaments is equivalent to that of a low-mass dwarf galaxy, of order a few times $10^{9}$\msol.
However, this mass is much less than the inferred mass of NCC~4696, so the infall does not represent a major perturbation to the elliptical galaxy. It is therefore a minor merger.}
\item{Imaging of the core of NGC~4696 shows evidence for at least two nuclear components in the galaxy \citep{Laine:2003p8500}.}
\item{The emission-line gas is characterized by a velocity dispersion of $\sim120\pm10$\kms, characteristic of being shock-excited.}
\item{As discussed in Section \ref{sec:kinematics}, the total emission-line luminosity from shocks is consistent with the estimated rate of dissipation of orbital energy in the filaments.}
\end{itemize}

In such a merger event, the dissipation of the orbital energy occurs through the drag forces generated by the motion of the merging galaxy gas through the hot galactic halo of the BCG. Shocks generated in the infalling gas would dissipate the orbital motion in this model, and therefore the signatures of shock excitation should be present. We investigate this possibility below.

% ------------------------------------------------------------------------------
\subsection{Radiative cloud shock models}\label{sec:shockmodels}

\begin{figure}
\epsscale{0.9}
\plotone{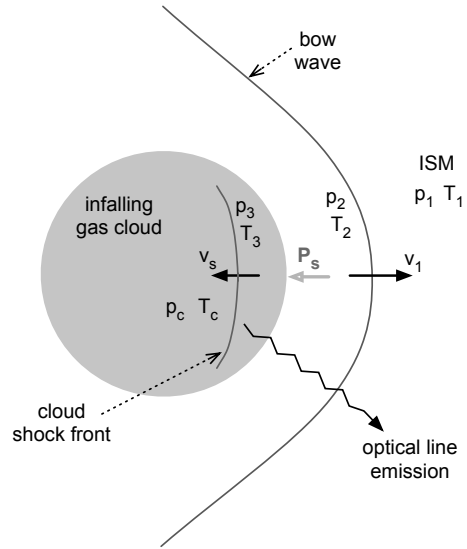}
\caption{Geometry of the infalling cloud/minor-merger shock model. The infall velocity of the cloud and therefore the bow wave velocity is $v_{1}$ and the cloud shock velocity is $v_{s}$. Other labelled parameters and components of the model are discussed in Section~\ref{sec:discussion}.}\label{fig:shockdiagram}
\end{figure}

We have run a grid of radiative shock models using the photoionization and shock modelling code MAPPINGS III, an updated version of the code described in \citet{Sutherland:1993p3718}. 
The motivation for these simulations is a model in which the filaments are comprised of clouds of cool gas that are remnants of a recently accreted, gas-rich, neighboring dwarf galaxy. These clouds are falling into the hot ISM halo of NGC~4696, which is compressed and heated by the ram pressure of the cloud's motion. The increased external pressure drives shocks into the cooler, dense cloud gas. 
The cooling time in the dense cloud material is short, so the cloud shocks are radiative and the observed optical line emission is produced in the cooling region behind the shocks. Figure~\ref{fig:shockdiagram} shows a simple schematic of this model and is further discussed in Section~\ref{sec:discussion}.

In the 1D plane-parallel shock models, the pre-shock medium is taken to be the warm atomic component of the infalling gas, rather than the dense molecular medium that is visible in the dust absorption. The reason for this choice is as follows. The total surface brightness of a radiative shock of velocity $v$ propagating into a medium of pre-shock density $\rho$ scales as the mechanical energy flux through the shock $S=1/2\rho v^3$. For a system of shocks propagating into a fractal medium with a wide range of temperatures,  the driving (ram) pressure  $P=\rho v^2 = $~constant. Thus $v \propto \rho^{-1/2}$ and therefore  $S \propto  \rho^{-1/2}$. High density regions do not significantly contribute to the total shock luminosity, not only because of this density-dependent factor, but also because the fractional filling factor of the denser cloud gas is small, reducing the total area that can radiate. Furthermore, in shocks propagating into molecular gas, the dominant luminosity is in collisionally-excited H$_2$ infrared emission, not in atomic line emission. Indeed infrared emission lines from H$_{2}$ gas are also seen, and their luminosity is correlated with the optical line emission \citep{Edge:2002p2508,Johnstone:2007p2172}, suggesting that these are produced by a common shock mechanism.

The ionization state of the material prior to processing by the shocks (the pre-ionization) and the shock velocity are treated as independent parameters in the models. The shock velocity is varied between 80 and 220\kms\, in intervals of 20\kms, and the pre-ionization from neutral (i.e. the fraction of ionized hydrogen is $n_{\rm H II}/n_{\rm H}= 0$) to fully ionized ($n_{\rm H II}/n_{\rm H} = 1$), with the fraction of hydrogen that is ionized in the pre-shock material increasing in steps of 0.1. A transverse magnetic field that is consistent with equipartition of thermal and magnetic field energy is assumed ($B=5\mu$G for $n=10$\,cm$^{-3}$; $B \propto n^{-1/2}$). No external radiation field was added. 

The gas abundances in Table~\ref{tab:abund} are taken from those presented by \citet{Asplund:2009p9806}, with typical dust depletion factors, since dust physics is not treated explicitly in these models. Observations of the Centaurus cluster show that the ICM gas abundance in the vicinity of the galaxy and filaments is  $Z/Z_{\odot}\sim2$ \citep{Graham:2006p2039,Sanders:2006p2045} and this overall metallicity is used in the models. However, we also find that an enhanced nitrogen abundance, by a factor of two, is required to better fit the observations. Nitrogen is made in the more massive AGB stars with hot-bottom burning ($5 < M/M_{\odot} < 8$), so possibly this nitrogen enhancement is associated with relatively less high-mass star formation and an older stellar population in the dense cluster environment.

\begin{table}
\centering
\caption{Solar abundances $Z_{\odot}$, the model abundance set $\log(n_{X}/n_{H})$, and included dust depletion factors $\log(D)$.\label{tab:abund}}
\begin{tabular}{@{}lccc@{}}
\tableline \tableline 
 Element &  $Z_{\odot}$  & Models  &  $\log(D)$ \\
\tableline
H	&	1.00		&	1.00		&	0.00		\\
He	&	-1.02	&	-0.92	&	0.00		\\
C	&	-3.57	&	-3.04	&	-0.30	\\
N	&	-4.17	&	-3.34	&	-0.20	\\
O	&	-3.31	&	-3.01	&	-0.23	\\
Ne	&	-4.07	&	-3.77	&	0.00		\\
Na	&	-5.75	&	-5.45	&	-0.60	\\
Mg	&	-4.43	&	-4.13	&	-1.15	\\
Al	&	-5.56	&	-5.36	&	-1.44	\\
Si	&	-4.49	&	-4.19	&	-0.88	\\
S	&	-4.86	&	-4.56	&	-0.34	\\
Cl	&	-6.63	&	-6.33	&	-0.30	\\
Ar	&	-5.60	&	-5.30	&	0.00		\\
Ca	&	-5.69	&	-5.39	&	-2.52	\\
Fe	&	-4.53	&	-4.23	&	-1.37	\\
Ni	&	-5.79	&	-5.49	&	-1.40	\\
\tableline
\end{tabular}
\end{table}

\begin{figure}\begin{center}
\epsscale{1}
\plotone{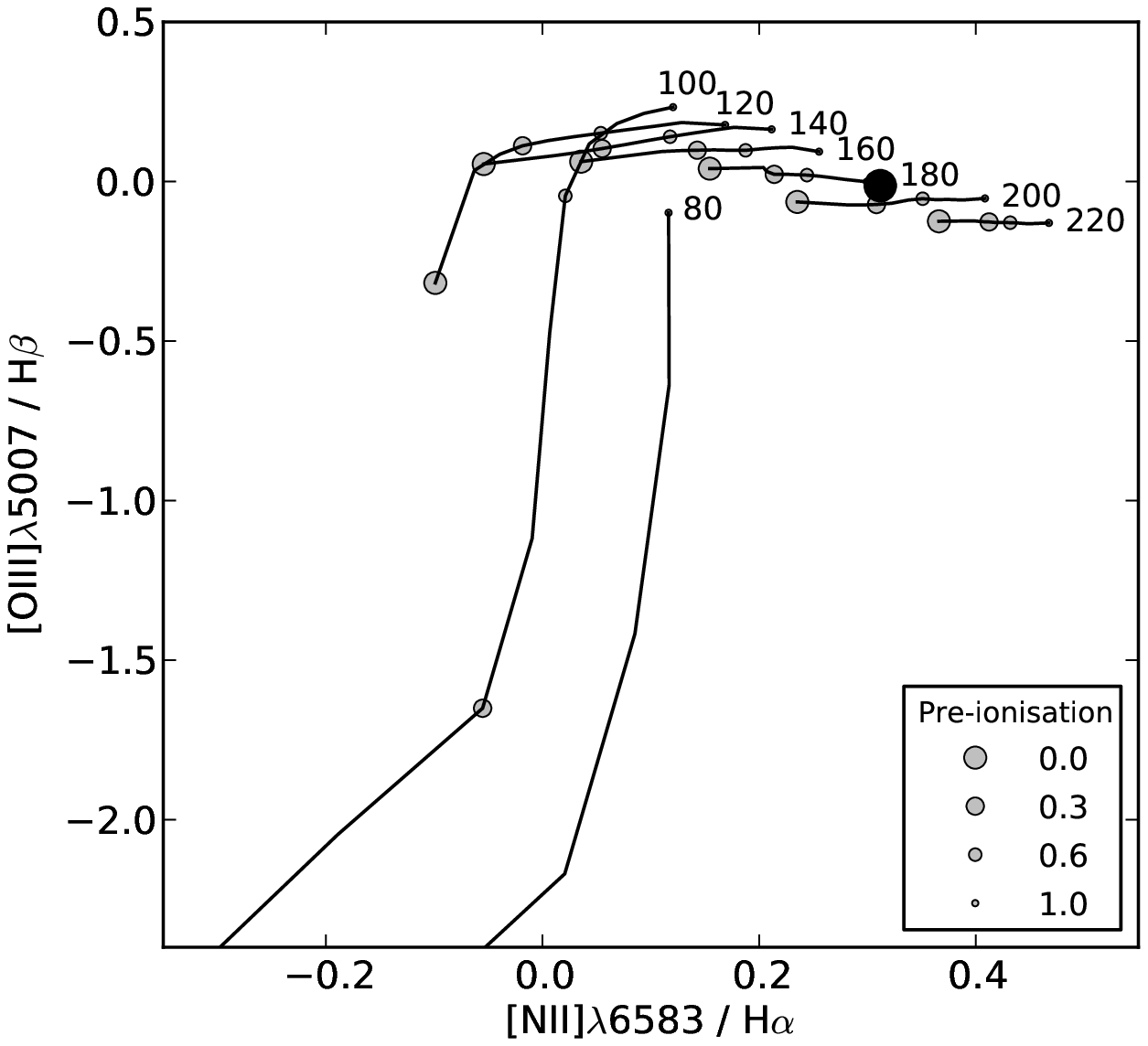}
\caption{Shock model results and measured line ratios for NGC~4696, plotted on the flux-ratio diagnostic diagram of [\ion{N}{2}] 6583\AA/H$\alpha$ \emph{vs.} [\ion{O}{3}] 5007\AA/H$\beta$. 
The solid lines connect the values from models of the same shock velocity, from 80\kms\, to 220\kms\, as labelled. The pre-ionization varies for the models that lie along these constant-velocity lines: the values corresponding to models with hydrogen pre-ionization fractions of 0, 0.3, 0.6, and 1.0 are marked by open circles that decrease in size with increasing ionization fraction.}
\label{fig:shockmodels}
\end{center}\end{figure}

\begin{figure}\begin{center}
\epsscale{1}
\plotone{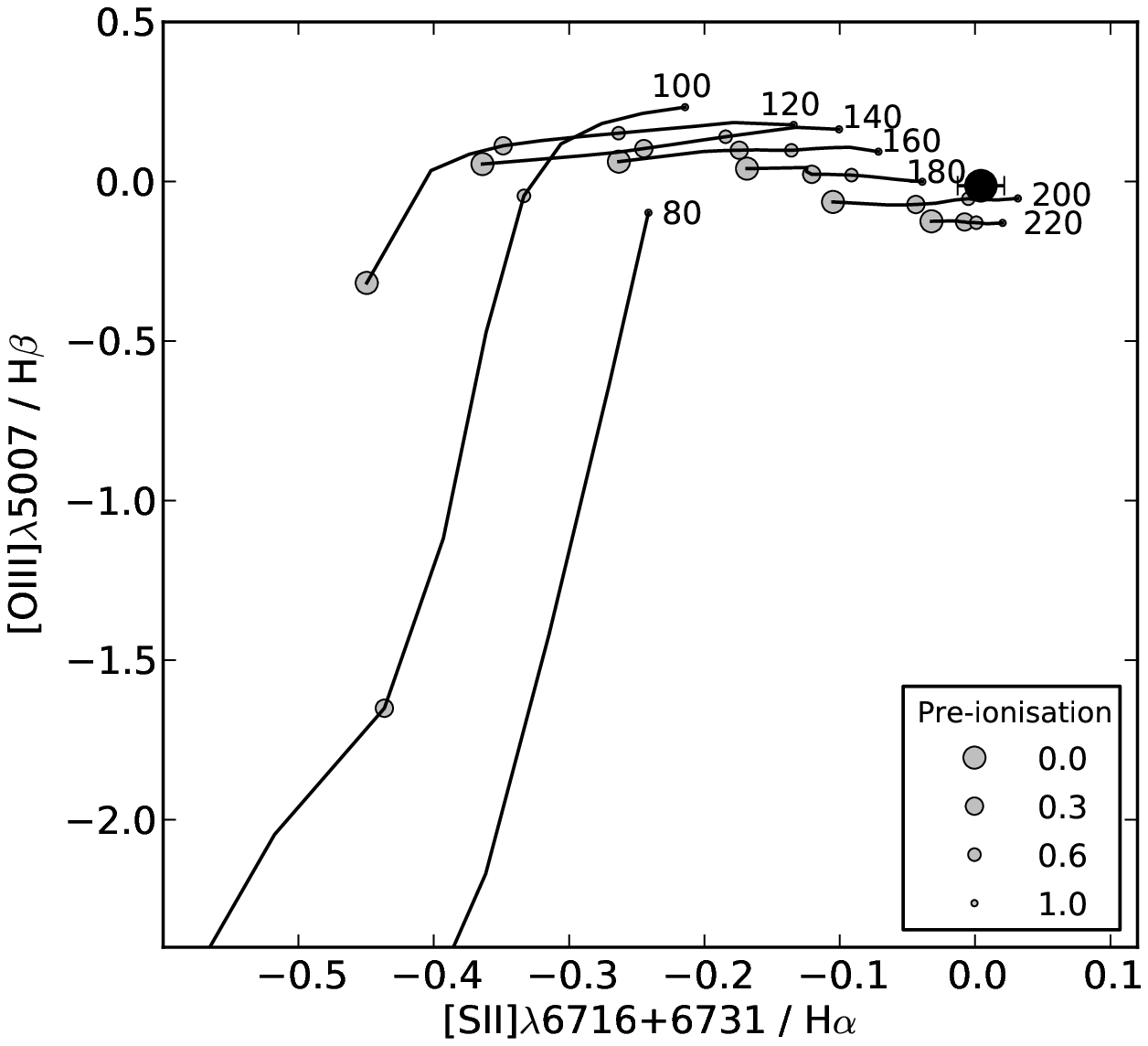}
\caption{Shock model results and measured line ratios for NGC~4696, plotted on the flux-ratio diagnostic diagram of [\ion{S}{2}] 6716+6731\AA/H$\alpha$ \emph{vs.} [\ion{O}{3}] 5007\AA/H$\beta$. Labels as in Figure~\ref{fig:shockmodels}.}\label{fig:shockmodels2}
\end{center}\end{figure}

\begin{figure}\begin{center}
\epsscale{1}
\plotone{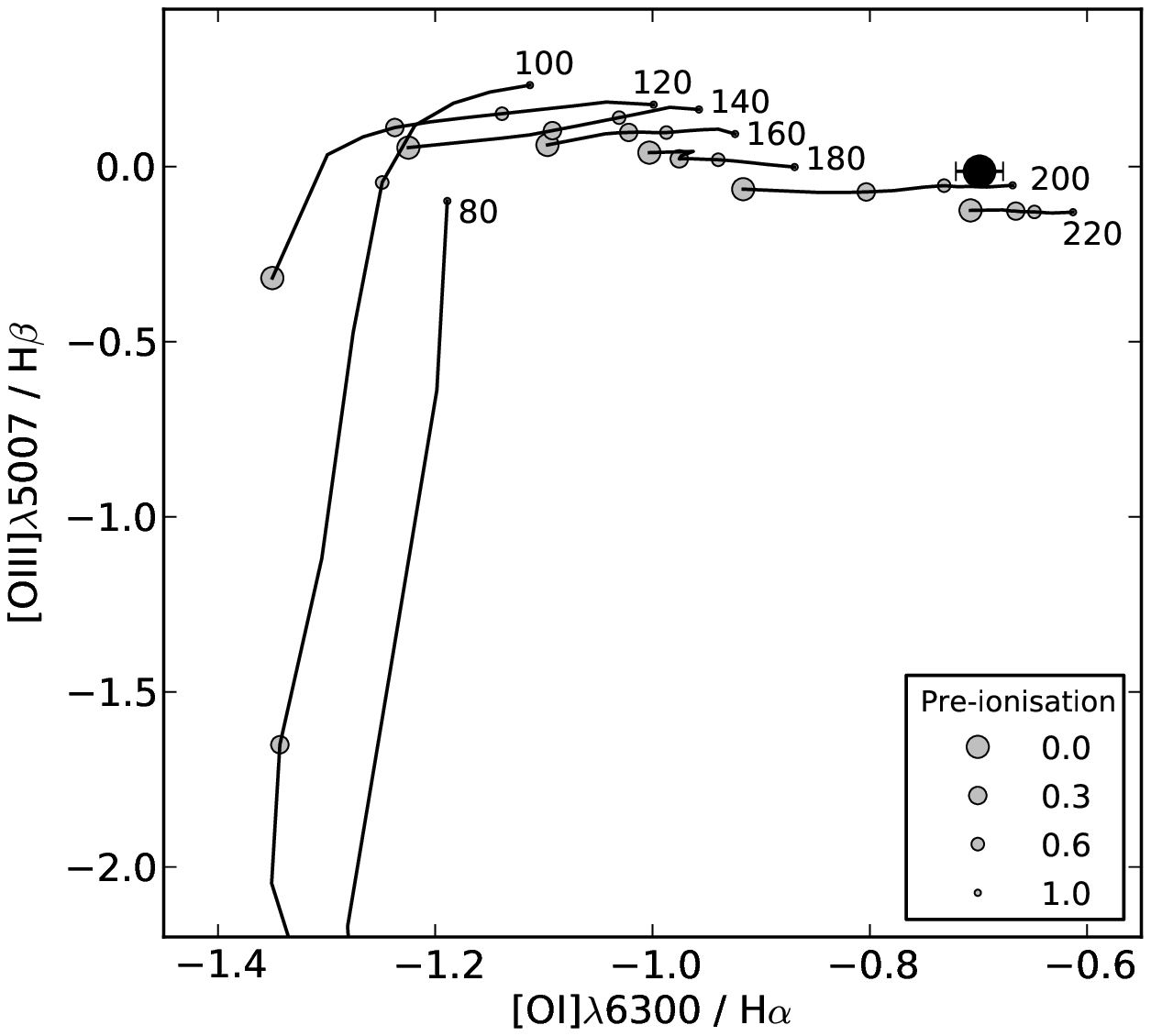}
\caption{Shock model results and measured line ratios for NGC~4696, plotted on the flux-ratio diagnostic diagram of [\ion{O}{1}] 6300\AA/H$\alpha$ \emph{vs.} [\ion{O}{3}] 5007\AA/H$\beta$. Labels as in Figure~\ref{fig:shockmodels}.}\label{fig:shockmodels3}
\end{center}\end{figure}

\begin{figure}\begin{center}
\epsscale{1}
\plotone{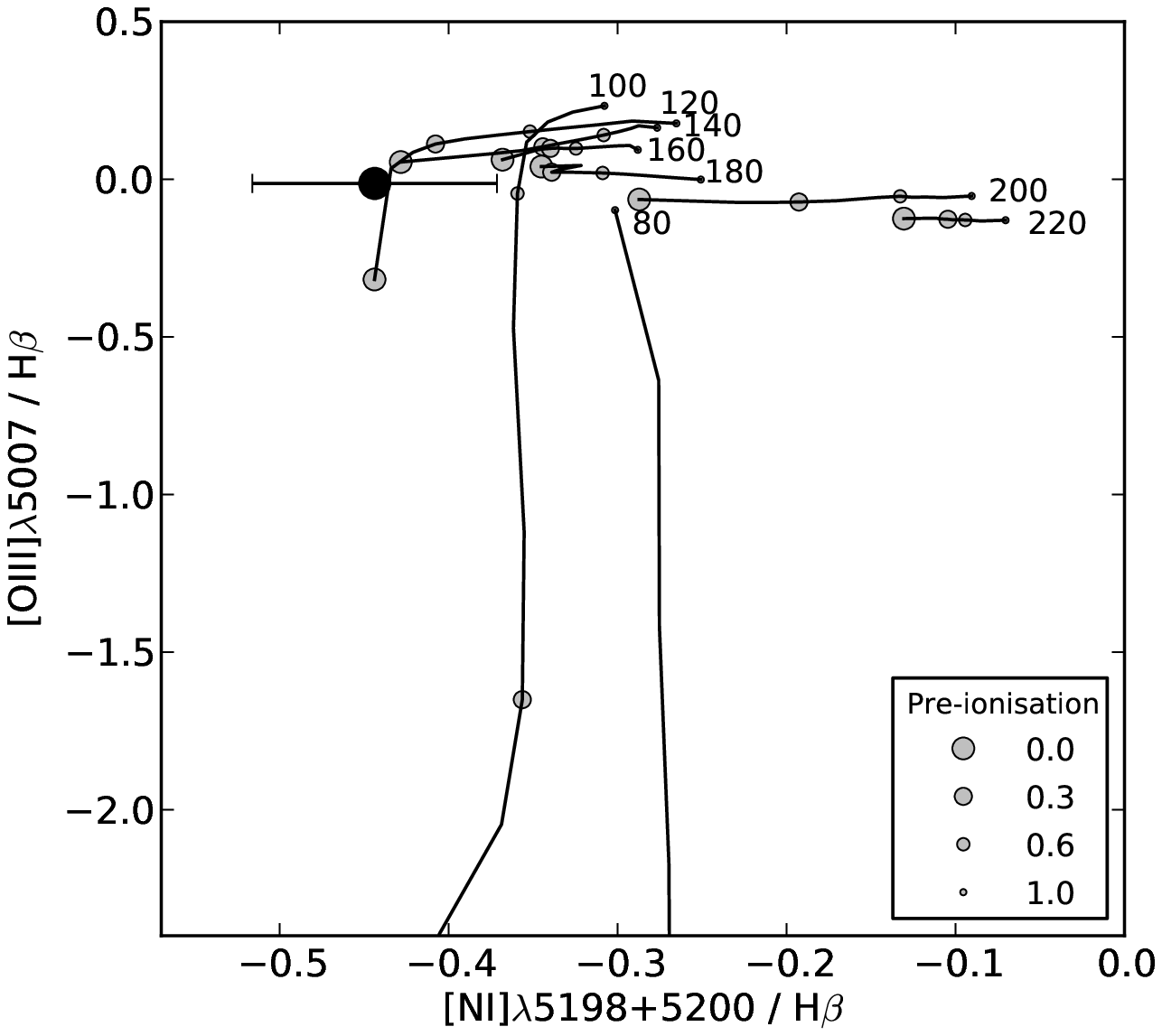}
\caption{Shock model results and measured line ratios for NGC~4696, plotted on the flux-ratio diagnostic diagram of [\ion{N}{1}] 5198+5200\AA/H$\beta$ \emph{vs.} [\ion{O}{3}] 5007\AA/H$\beta$. Labels as in Figure~\ref{fig:shockmodels}.}\label{fig:shockmodels4}
\end{center}\end{figure}

\begin{figure}\begin{center}
\epsscale{1}
\plotone{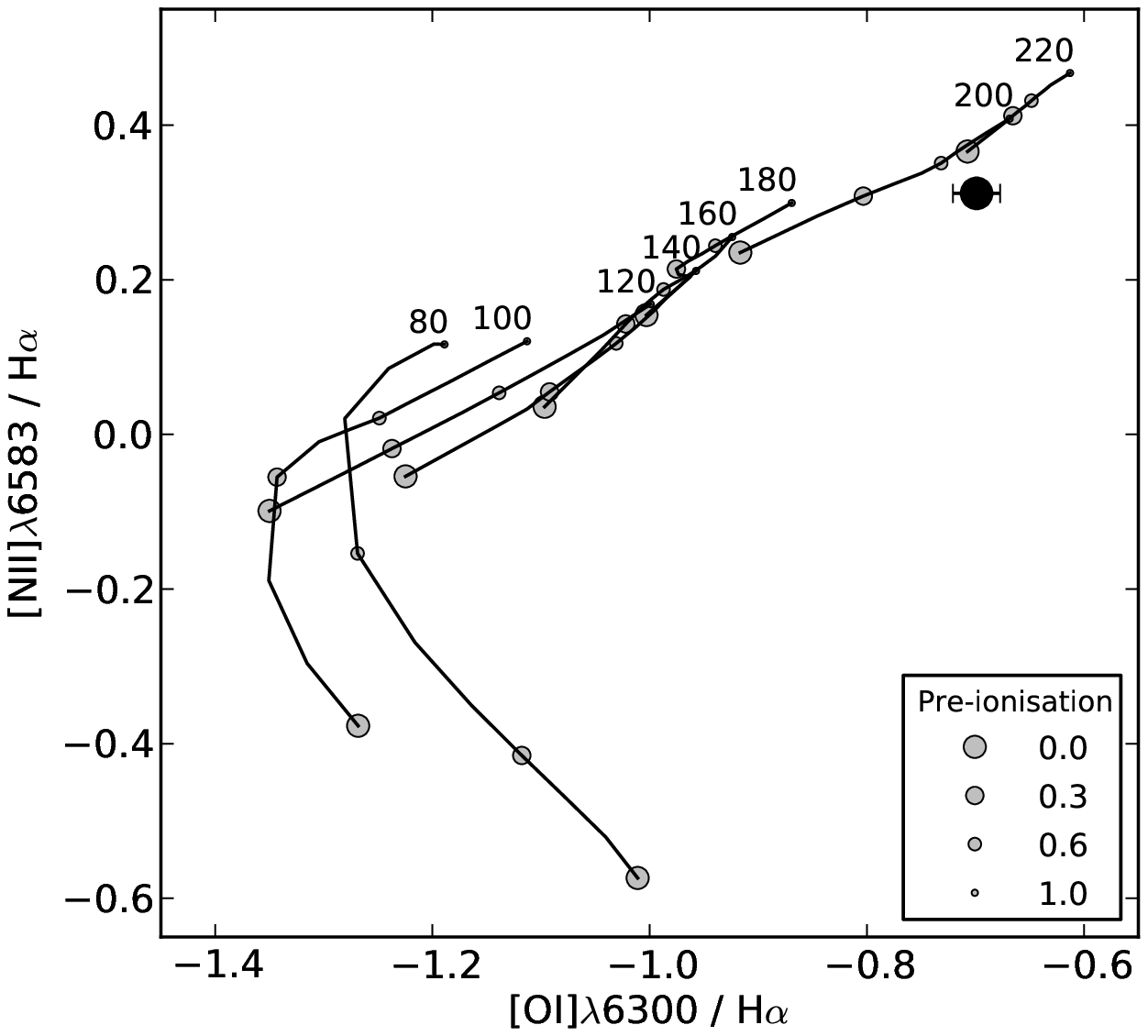}
\caption{Shock model results and measured line ratios for NGC~4696, plotted on the flux-ratio diagnostic diagram of [\ion{O}{1}]\,6300\AA/H$\alpha$ \emph{vs.} [\ion{N}{2}]\,6583\AA/H$\alpha$. Labels as in Figure~\ref{fig:shockmodels}.}\label{fig:shockmodels5}
\end{center}\end{figure}

\begin{figure}\begin{center}
\epsscale{1}
\plotone{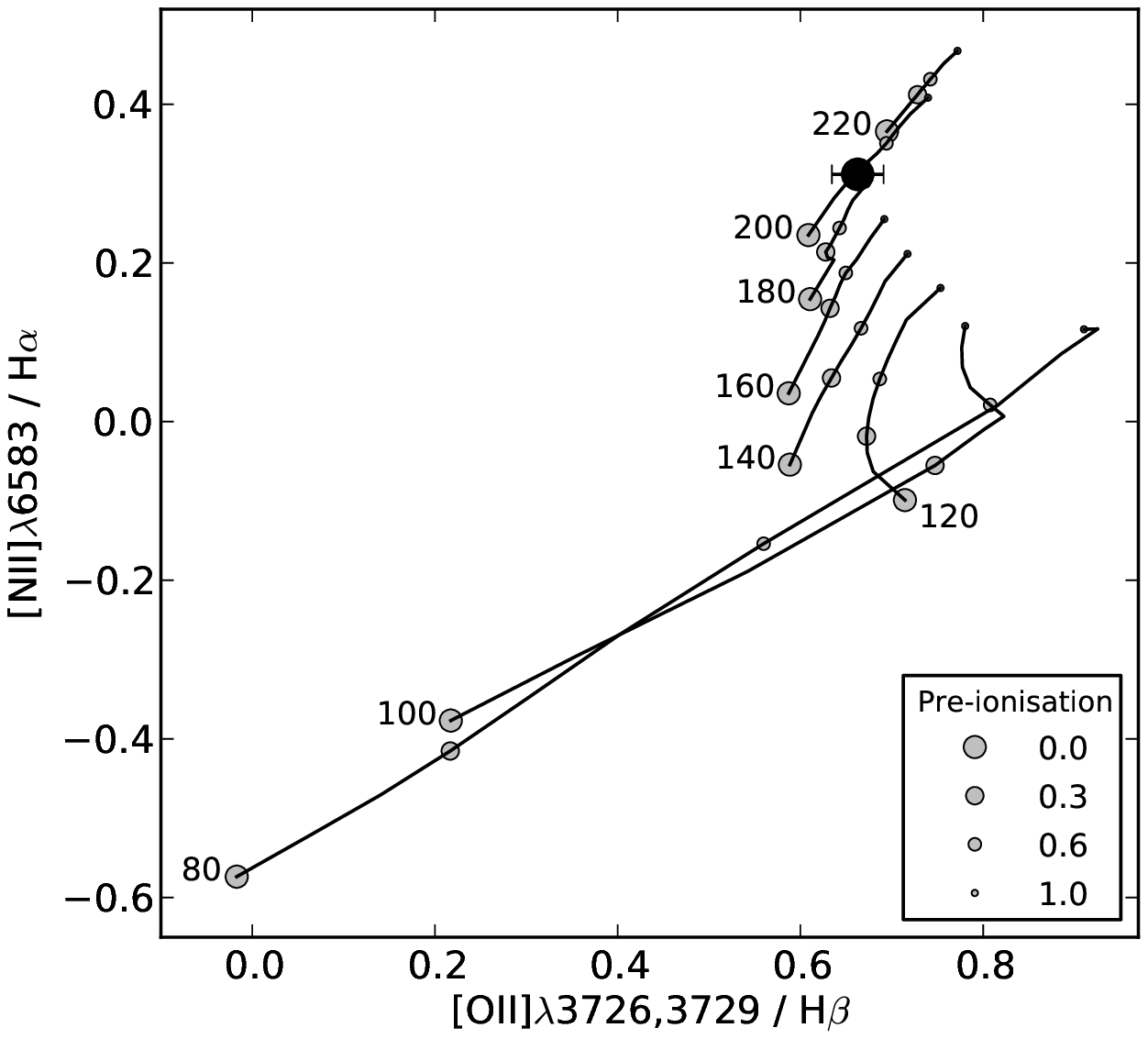}
\caption{Shock model results and measured line ratios for NGC~4696, plotted on the flux-ratio diagnostic diagram of [\ion{O}{2}]\,3727\AA/H$\beta$ \emph{vs.} [\ion{N}{2}]\,6583\AA/H$\alpha$. Labels as in Figure~\ref{fig:shockmodels}.}\label{fig:shockmodels6}
\end{center}\end{figure}

With this abundance set, the shock model is sought that provides a least-squares best fit to the values of the following line flux ratios:  [\ion{O}{3}]\,$\lambda$5007/H$\beta$; [\ion{N}{1}]\,$\lambda$5198+5200/H$\beta$;  [\ion{O}{1}]\,$\lambda$6300/H$\alpha$; [\ion{N}{2}]\,$\lambda$6583/H$\alpha$; and [\ion{S}{2}]\,$\lambda$6731/H$\alpha$. 
The individual line ratios from the grid of models are shown on the line flux ratio diagrams in Figures~\ref{fig:shockmodels} to \ref{fig:shockmodels6}, along with the line ratios measured in the filaments of NGC~4696. The range of parameter space plotted here is much narrower than the normal range displayed on these type of diagnostic diagrams. 
All except the [N~I] $\lambda$5198+5200/H$\beta$ ratio are consistent with shock models with velocities in the range $v_s = 180-200$km~s$^{-1}$. 
We have determined the best-fit model spectrum by minimizing the RMS fit to all these ratios. The resulting best-fit model has a shock velocity $v_s = 180$\,\kms. The best-fit model also has complete pre-ionization of H, though it can be seen from the diagnostic diagrams in Figures~\ref{fig:shockmodels} to \ref{fig:shockmodels6} that the results are not very sensitive to the pre-ionization of the high-velocity models. However, at shock velocities of 100-200\kms\, the pre-shock gas will be photoionized by the upstream radiation flux from the shocked gas, so the most highly-ionized model is appropriate for the these velocities. 
The best-fit shock spectrum (at optical wavelengths) is given in Table~\ref{tab:shockspec}. 


\begin{table}
\centering
\caption{The best-fit model shock spectrum \label{tab:shockspec}}
\begin{tabular}{@{}lccp{0.5cm}lcc@{}}
\cline{1-3}\cline{5-7}	 
{$\lambda$(\AA)} & $F/F_{\rm H\beta}$ & Ion & & {$\lambda$(\AA)}	& $F/F_{\rm H\beta}$ & Ion \\
\cline{1-3}\cline{5-7}
3425.81	&	0.2341	&	Ne V	&	&	5015.70	&	0.0286	&	He I	\\
3726.03	&	2.0440	&	O II	&	&	5197.82	&	0.2308	&	N I	\\
3728.73	&	2.4180	&	O II	&	&	5200.17	&	0.2854	&	N I	\\
3750.15	&	0.0300	&	H I	&	&	5754.50	&	0.0596	&	N II	\\
3770.63	&	0.0390	&	H I	&	&	5875.60	&	0.1106	&	He I	\\
3797.90	&	0.0521	&	H I	&	&	6300.20	&	0.3445	&	O I	\\
3835.38	&	0.0718	&	H I	&	&	6363.67	&	0.1135	&	O I	\\
3868.62	&	0.3361	&	Ne III	&	&	6547.96	&	1.8400	&	N II	\\
3888.60	&	0.0791	&	He I	&	&	6562.80	&	2.9870	&	H I	\\
3889.05	&	0.1032	&	H I	&	&	6583.34	&	5.4180	&	N II	\\
3967.34	&	0.1012	&	Ne III	&	&	6678.20	&	0.0314	&	He I	\\
3970.07	&	0.1560	&	H I	&	&	6716.31	&	1.3960	&	S II	\\
4026.20	&	0.1757	&	He I	&	&	6730.68	&	1.1790	&	S II	\\
4068.50	&	0.0694	&	S II	&	&	7065.20	&	0.0111	&	He I	\\
4076.27	&	0.0229	&	S II	&	&	7135.67	&	0.0571	&	Ar III	\\
4101.73	&	0.2539	&	H I	&	&	7319.65	&	0.0201	&	O II	\\
4340.46	&	0.4614	&	H I	&	&	7320.24	&	0.0620	&	O II	\\
4363.15	&	0.0983	&	O III	&	&	7330.10	&	0.0333	&	O II	\\
4471.50	&	0.0378	&	He I	&	&	7330.70	&	0.0325	&	O II	\\
4566.78	&	0.1473	&	Mg I	&	&	7750.95	&	0.0137	&	Ar III	\\
4685.74	&	0.0628	&	He II	&	&	8617.00	&	0.0142	&	Fe II	\\
4714.10	&	0.0133	&	Ne IV	&	&	8727.66	&	0.0149	&	C I	\\
4724.11	&	0.0144	&	Ne IV	&	&	8862.79	&	0.0142	&	H I	\\
4725.60	&	0.0133	&	Ne IV	&	&	9014.91	&	0.0189	&	H I	\\
4861.32	&	1.0000	&	H I	&	&	9068.44	&	0.0956	&	S III	\\
4958.83	&	0.3342	&	O III	&	&	9229.02	&	0.0262	&	H I	\\
5006.77	&	0.9627	&	O III	&	&	9530.44	&	0.2370	&	S III	\\
\cline{1-3}\cline{5-7}	
\end{tabular}
\end{table}

Insufficient observed strength of the [\ion{O}{3}]~$\lambda$4363\,\AA\, line has been used to argue against shock models as a possible exciting mechanism in the filaments of BCGs \citep[e.g.][]{Sabra:2000p2757,Voit:1997p2496}.
We do not detect the [\ion{O}{3}] line in the spectra of NGC~4696, however this is consistent with the predictions of the shock models described here. We can place an upper limit on the line strength of $\lesssim\scinot{6}{-17}$\surfb\, and therefore the ratio of fluxes [\ion{O}{3}]/H$\beta\lesssim0.35$, at the emission-line peak. 
From the model spectrum in Table~\ref{tab:shockspec}, the expected [\ion{O}{3}]~$\lambda$4364\,\AA\,flux is approximately 10\% of the H$\beta$ line strength, and so below the detectable level and consistent with the data.

The only other emission line that is produced in the shock model spectrum at a level that may be detectable in our data is [\ion{Ne}{3}]\,$\lambda$3869\,\AA; [Ne III]/H$\beta$ = 0.34 from Table~\ref{tab:shockspec}. However, as indicated above, at the blue end of the spectrum the ratio of detectable line flux to the flux in H$\beta$ at the line-emission peak is $\gtrsim0.35$, so the predicted [\ion{Ne}{3}] strength lies at the edge of our sensitivity limit. There is evidence for a marginal detection of this line at the emission peak (see Figure~\ref{fig:contsub}), with an integrated flux of $\approx$\scinot{(8\pm5)}{-16}\flux\, (after extinction correction), but the fluxes in individual pixels are below the limit of detection.

Though consistent in other measures, the shock models over-predict the strength of the [\ion{N}{1}]\, $\lambda$5200\AA\, doublet by a factor of $\sim$1.4, the effect of which is seen in Figure~\ref{fig:shockmodels4}. Measurement errors may contribute to this discrepancy: the [\ion{N}{1}] line flux is close to the detection limit in the blue spectra, and the line lies on the wing of the deep \ion{Mg}{1} absorption feature, which could introduce systematic errors in the galaxy continuum subtraction and line fitting that are not represented in the uncertainties given in Table~\ref{tab:fluxes}.
Additionally, in the models the flux of the [\ion{N}{1}] line is very sensitive to the boundary conditions that  terminate the simulations and so is strongly affected by the selected model configuration. 

In some sense, depending on projection effects, the velocity dispersion from the emission-line widths reflects the shock velocity. The line-flux ratios also depend on the shock velocity and should therefore correlate with the velocity dispersion. So if the line ratios are uniform, then the velocity dispersion is also expected to be uniform. This is indeed observed to be the case in the main filament.

We conclude then, that radiative shock models having super-solar abundances, and shock velocities of around $v_s = 180$\,\kms\, provide a good fit to the observed emission-line spectrum of NGC~4696 across the main filament. 
We will now examine how we can use this cloud-shock model to further constrain the parameters of the infalling emission-line gas. 

%-------------------------------------------------------------------------------
\subsection{Global model parameters} \label{sec:discussion}

In our model, the infalling motion of the gas clouds generates the ram pressure that drives radiative shocks into the cloud material. The cloud velocity and the properties of the hot, ionized gas in the galaxy halo (ISM) determine the magnitude of the driving pressure that is obtained. 

Figure~\ref{fig:shockdiagram} outlines the geometry of the model and the components referred to in the following discussion. As a cool gas cloud falls through the ambient medium, a bow wave, or shock if supersonic, forms in the ISM ahead of the cloud. The properties of the ambient ISM gas are denoted with the subscript 1, and $v_{1}$ is the velocity of the ISM relative to the cloud (i.e. having magnitude equal to the infall velocity). Immediately behind the bow wave the gas has pressure, density, and temperature $P_{2}$,  $\rho_{2}$,  $T_{2}$, and the velocity relative to the bow wave is $v_{2}$. 
If $v_{1}$ is greater than the sound speed in the ISM $c_{s,1} = (\gamma\,p_{1}/\rho_{1})^{1/2}$, the bow wave is a shock, and the Mach number is $M_{1} = v_{1}/c_{s,1} > 1$. In this case the pressure will increase across the shock front. If the motion is subsonic $v_{1}<c_{s,1}$, so $M_{1}<1$, and the pressure is constant across the bow wave $p_{2}=p_{1}$.

We determine the maximum pressure that acts on the cloud (due to its motion through the ISM) as the stagnation pressure at the leading surface of the cloud where the flow comes to rest. The Bernoulli equation along a streamline traversing the post-bow-wave region, assuming an adiabatic, compressible flow behind the bow wave, gives:
\[  \frac {1}{2} v_{s}^2 + h_{s} = \hbox{Streamline Constant} = \frac {1}{2} v_2^2 + h_2 \]
where $v_{s}=0$ is the velocity at the stagnation point, and $h$ is the specific enthalpy,
\[ h = \frac{\gamma}{\gamma-1}\frac{p}{\rho} = \frac{c_{s}^2}{\gamma-1}. \]
From this, the stagnation pressure can be expressed as
\begin{equation}
p_s = p_2 \, \left[ 1 + \frac {\gamma-1}{2} M_2^2 \right]^{\gamma/(\gamma-1)}
\label{eq:pstag}
\end{equation}
in terms of the conditions in the flow behind the bow wave. We compare the infall velocity of the cloud to the ambient sound speed in the ISM to estimate the Mach number of the bow wave ($M_{1}$) and hence the properties of the gas behind it.

\cite{Graham:2006p2039} present density and temperature profiles of the IGM gas derived from X-ray observations obtained with Chandra, XMM-Newton, and ROSAT. We use these to estimate the pressure and sound speed in the halo of NGC~4696 and constrain the pressure driving the shocks in the filament clouds. 
At a radius of $\sim1$\,kpc (5\arcsec), which corresponds to the inner region of the filaments, the X-ray-emitting plasma has a temperature of $T_{1}=\scinot{(8\pm3)}{6}$\,K and hydrogen particle density $n_{H,1}=0.13\pm0.04$\pcc. The sound speed in this gas is then $c_{s,1}=440\pm50$\kms\, and the pressure is $p_{1}/k = \scinot{(2.5\pm0.5)}{6}$\,K\pcc.

From our IFU observations of the inner filament, we assume an infall velocity of the clouds in the filament of $v_{1}\sim 300$\kms, approximately the maximum line-of-sight velocity of the emission in the filament.
The Mach number of the flow is then $M_{1}\sim 0.7$. In this case the flow is in the transonic regime and a weak bow shock forms. 
For a transonic shock $M_{1}\approx 1 \approx M_{2}$, and the pressure remains approximately constant across the shock: $p_{1}\approx p_{2} \approx \rho_{1}\,v_{1}^{2}/\gamma$. From Equation~\ref{eq:pstag}, the expression for $p_{s}$ becomes
\begin{eqnarray}
p_s &\approx& \frac {\rho_1 v_1^2}{\gamma} \left[\frac {\gamma+1}{2} \right]^{\gamma/(\gamma-1)} \\
&\approx& 1.23 \, \rho_1 v_1^2 = 2.05 \, p_1 \quad \hbox{for} \quad \gamma = \frac {5}{3},
\end{eqnarray}
so the static pressure of the X-ray-emitting ISM is enhanced by a factor of approximately two at the surface of the cloud.

At the radius of the inner filaments, the mass density of the ISM is $\rho_{1}=1.4\,m_{H}\,n_{H,1} \sim \scinot{3.0}{-25}$\,cm$^{-3}$. With a bow-shock velocity of $v_{1}\sim300$\kms, the resulting stagnation pressure is $p_{s}/k = \scinot{2.4}{6}$\,K\pcc. 
This post-bow-shock pressure drives the cloud shock by providing an equivalent pressure to the gas behind the cloud shock. We assume this pressure remains approximately constant across the post-cloud-shock region of the cloud (denoted by subscript 3 in Figure~\ref{fig:shockdiagram}).

For a strong shock, the pressure in the post-cloud-shock region $p_{3}$ is related to the pre-shock cloud gas properties ($p_{c}$, $\rho_{c}$) as follows:
\begin{eqnarray}
 p_{3}  &=& \frac{2}{\gamma+1}\rho_{c}v_{s}^{2} \\
           &=& \frac{2}{\gamma+1} \mu\,m_{H}\, n_{H,c}\, v_{s}^{2}
\end{eqnarray}
and the pre-shock hydrogen particle density in the cloud is then
\begin{eqnarray}
n_{H,c}  &\approx& \frac{\gamma+1}{2} \frac{p_{s}}{ \mu\,m_{H}\,v_{s}^{2} } \\
              &\approx& \scinot{1.76}{9}\, p_{s} 
\label{eq:nHbow}
\end{eqnarray}
for $\gamma = 5/3$ and the best-fit shock-model velocity $v_{s}=180$\kms. Hydrogen is fully ionized in the pre-shock cloud gas, so $\mu = 1.4$.
From these results we can estimate the density of the ionized cloud material. For the inner filaments we then find a hydrogen density $n_{H,c} \approx 1$\,cm$^{-3}$.

The pre-shock density in these models cannot be constrained by the [\ion{S}{2}]\,$\lambda$6716/6731\,\AA\,ratio, as the observations are consistent with the low density limit ($n_e < 300$\pcc\, in the recombination zone of the shock, with $T_e \sim 6000$\,K). This prevents a direct measurement of the ram pressure associated with the shock. A somewhat less accurate estimate can be obtained from the H$\alpha$ surface brightness, since this scales as the total emitted shock surface luminosity. 
The best-fit shock model has emitted surface luminosity in H$\alpha$:  
\begin{equation}
S_{\rm H\alpha} = \frac{S_{\rm H\alpha}}{S_{\rm total}} \rho v^{3} = 9.6 \times 10^{-5} \left[ \frac{n_{e,c}}{{\rm cm^{-3}}} \right]\hbox{\,\flux.}\label{eq:shockSB}
\end{equation}
where $n_{e,c}$ is the electron density in the pre-shock cloud material.

The average H$\alpha$ surface brightness in the main filament measured in the 3\arcsec-diameter aperture around the region of the line peak is $I_{\rm H\alpha}\sim\scinot{9.4}{-16}$\surfb. The measured surface brightness varies across the filaments, with a typical value in the outer region of the main filament of $\sim\scinot{1}{-16}$\surfb. This corresponds to a measured surface luminosity from the inner filament of 
\begin{equation}
S_{H\alpha} = I_{H\alpha}~\frac{4\pi D^2}{\phi}\left(\frac{206264.8}{D}\right)^2 = \frac{5.35\times10^{11}I_{H\alpha}}{\phi}
\end{equation}
where $\phi$ is the filling fraction of shock emission inside the 3\arcsec\, aperture. This fraction is highly uncertain. However, if we assume $\phi=1$, we obtain $S_{\rm H\alpha} \sim \scinot{5.2}{-4}$\flux\, in the central filaments and $\sim\scinot{5.5}{-5}$\flux\, in the outer region. Using the relation between the density and surface luminosity in Equation~\ref{eq:shockSB}, we find a pre-shock density, $n_{e,c} \sim 5$\pcc~near the nucleus, and $n_{e,c} \sim 0.6$\,\pcc~in the outer regions. 
In the ionized pre-shock medium, the equivalent hydrogen densities are $\sim4$\,cm$^{-3}$ and $\sim0.5$\,cm$^{-3}$. 
There are uncertainties associated with these estimates, most notably in the filling factor of the shock surfaces in the surface luminosity estimates. However, the result is consistent with the density estimates of $\sim1$\,cm$^{-3}$ derived through the bow-shock/cloud-shock pressure arguments presented previously. 
The radiative shock model appropriate to the nuclear region with $v_s = 180$\kms, $n_{\rm 0} \sim 4$\pcc, and equipartition magnetic field $B = 5\,\mu$G produces a mean density in the [\ion{S}{2}]-emitting zone of the shock of 190\pcc, consistent with the [\ion{S}{2}] $\lambda$6716\AA/6731\AA\, flux ratio being observed at its low-density limit. 
These results provide support for the feasibility of the model as a mechanism for generating shock-excited emission-line spectra in the extended filaments of NGC~4696.

The increased pressure in the bow shock region will lead to an enhancement of the X-ray emissivity in the vicinity of filaments. This can provide a natural explanation of the correlation between the X-ray and the optical emission observed by \citet{Crawford:2005p334}. These authors measure the $0.1-10$\,keV flux in the filament 22\arcsec\,to the south-east of the nucleus to be $\sim\scinot{7}{-15}$\surfb. From our data, the H$\alpha$ surface brightness in this outer filament is $\sim\scinot{5}{-17}$\surfb, so the H$\alpha$ flux in the region is approximately 0.7\% of the soft X-ray flux.
The total surface brightness in the cloud shocks, from the model, is about 150 times the H$\alpha$ surface brightness. Combining this with the measured ratio of X-ray to H$\alpha$ flux, the ratio of the X-ray to total cloud shock surface brightness is estimated to be ${S_{X}}/{S_{c}} \sim 0.9$. Thus the energy dissipation in the X-ray plasma is comparable to that of the cloud shocks. This ratio might be lower if there are faster (non-radiative) shocks in the infalling gas, as these could provide soft X-ray luminosity associated with the infalling gas, rather than the shocked ISM. Nonetheless, it is clear that the gas-dynamic drag of the hot ISM is an important factor in determining the accretion timescale of the infalling gas towards the nucleus.

%--------------------------------------------------------------------------------------------------
\subsection{Extended LINER emission and shocks in other galactic environments}

Although LINER emission in galaxies is frequently a nuclear phenomenon, there are a range of galactic environments that host extended, extra-nuclear, LINER-like emission regions. Other than cool-core cluster BCGs, these include galactic winds, AGN jets, and interacting or merging galaxies. 

A link between LINER-like emission and shock excitation was proposed in early studies of LINER galaxies and has since been extensively discussed and debated in the literature \citep[e.g.][]{Dopita:1994p1099}. 
Indeed, from early investigations of LINER spectra, \citet{Koski:1976p11034}, \citet{Fosbury:1978p9939}, and \citet{Heckman:1980p10009} concluded that shock heating is responsible for producing the low-ionization spectrum in LINERs. More recently, \citet{Filippenko:2003p2634} noted that despite a lack of strong evidence for shock heating in galactic nuclei, shocks are likely to dominate in cases of spatially-extended LINER-like ionization.  A connection between LINER emission and shock excitation has been proposed in models of galaxy `superwinds', e.g. in NGC~839 \citep{Rich:2010p10874} and NGC~3079 \citep{Filippenko:1992p11039}, and in interactions and mergers \citep[e.g.][]{GonzalezDelgado:1996p11040}.

Most interesting in the context of this paper, are recent findings that increasingly point to shock excitation of LINER-like emission in the tidal regions of galaxy mergers and interactions. 
\citet{Ogle:2007p10860} present Spitzer observations of the luminous infrared emission associated with the double galaxy system 3C 326. The northern galaxy of the pair produces strong H$_{2}$ line emission and a LINER-type optical spectrum. The authors investigate several heating mechanisms for the IR emission and conclude that the molecular hydrogen spectrum may be powered by accretion shocks that arise in inflows induced by tidal interactions between the two galaxies. 
\citet{MonrealIbero:2010p10847,MonrealIbero:2006p10845} investigate the ionization of the extended emission regions in a sample of luminous infrared galaxies (LIRGs) using integral-field spectroscopy. They find that the emission in isolated systems is generally consistent with ionization by radiation from young stars. However, large fractions of the emission spectra from the extended regions in interacting pairs and more advanced mergers are consistent with ionization by shocks with velocities of $150-500$\kms.
In their model, the shocks are generated by tidally-induced, high-velocity flows that can be a result of the merging process \citep[e.g.][]{McDowell:2003p11080}. They suggest that tidal forces may be central to the origin of shocks that ionize the gas in the extended regions and produce the low-ionization spectrum.
%%% referee (10)
The most luminous BCGs do produce IR luminosities that place them in the class of LIRGs \citep[$L_{\rm IR} > 10^{11}$\,L$_{\Sol}$; e.g.][]{Egami:2006p3830}. While NGC~4696 is not such an IR-luminous system: $L_{\rm IR} = \scinot{2.3}{9}$\,L$_{\Sol}$ \citep{Kaneda:2007p8744}, results like those discussed above support the suggestion that infall and merger activity provides a mechanism for generating galaxy-wide LINER-like emission through intermediate-velocity shock excitation.

%%%referee (12)
Lastly, we note the observations presented by \citet{Edwards:2009p11292} of the BCG system in the Ophiuchus cluster, which make an interesting comparison with our results for NGC~4696. In Ophiuchus, the BCG does not produce detectable line emission, but is accompanied by a nearby extended object that emits strong low-ionization optical lines. This object is $\sim1.7$\,kpc from the center of the BCG on the sky and is separated by 600\kms\, in radial velocity. The red emission line ratios fall in the AGN or LINER regions of the diagnostic diagrams, but the [\ion{O}{3}]/H$\beta$ ratio is not available to distinguish between these classes. The authors conclude that it is most likely that the emission is produced within a large gas cloud falling towards the BCG through the galaxy halo. It is possible that the emission in this structure is excited by the ram pressure of its motion through the hot atmosphere of the BCG, as in the model we have discussed here. The Ophiuchus system may be in the early stages of a process of minor merger like the one that has occurred in NGC~4696. Further study is needed to better understand the nature of the interaction.

%------------------------------------------------------------------------------------------------------------------------------
\section{Conclusion} \label{sec:conclusion}
Our IFU observation of the extended emission surrounding NGC 4696 provides a picture of material falling in along the bright curved filament that encircles the core of the galaxy, with an emission-line spectrum - and hence excitation mechanism - that exhibits uniform properties over the inner filaments.

The kinematics indicate `orbital' trajectories, spiralling in toward the galaxy nucleus. In addition to other observational data that provide evidence for a recent merger event in the system, this prompts us to investigate a model in which the extended emission-line filaments form from the stripping of material during a minor merger event that also provides the source of energy that excites the line emission.
In this model, the infalling clouds in the filaments are remnants of a low-mass, gas-rich (containing approximately $10^{9}$\msol\,gas) cluster dwarf galaxy captured in the potential of the BCG. The motion of the filaments through the hot ICM produces bow waves that drive shocks into the clouds themselves. We present MAPPINGS models of the cooling behind these shocks and find that the optical emission-line spectrum is well-matched by $\sim$180\kms\, shocks propagating into ionized gas.

The minor-merger scenario described here has the benefit of naturally accounting for the high densities and pressures in the filaments relative to the surrounding medium (unlike direct cooling models that assume that the filaments are in pressure equilibrium with the environment). It also explains the kinematic properties of the filaments, and is energetically self-consistent. Furthermore, the presence of large quantities of dust closely associated with the optical filaments is consistent with the model of an infalling satellite. 

Excitation by low-velocity shocks is a mechanism that may be capable of explaining the underlying LINER-like properties of the extended filament emission in central cluster galaxies. Although such shocks may not be always driven by infall or mergers, there is an interesting parallel between the properties of the filaments in NGC~4696 and those of emission regions in LIRGS, which have been attributed to tidally-induced accretion shocks.

%------------------------------------------------------------------------------------------------------------------------------
\section*{Acknowledgments}
MD acknowledges the support from the Australian Department of Science and Education (DEST) Systemic Infrastructure Initiative grant and from an Australian Research Council (ARC) Large Equipment Infrastructure Fund (LIEF) grant LE0775546, which together made possible the construction of the WiFeS instrument. We would also like to thank the Australian Research Council (ARC) for support under Discovery projects DP0984657 (MD),  DP0664434 (MD and GB), and DP0342844 (PM and CF).
CF is grateful to Ralph Sutherland for helpful discussions on shock modelling.
This research has made use of the NASA/IPAC Extragalactic Database (NED) which is operated by  the Jet Propulsion Laboratory, California Institute of Technology, under contract with the National  Aeronautics and Space Administration.  
This research has also made use of NASA's Astrophysics Data System, of SAOImage DS9 \citep{Joye:2003p10689}, developed by the Smithsonian Astrophysical Observatory, and of the Hubble Legacy Archive, which is a collaboration between the Space Telescope Science Institute (STScI/NASA), the Space Telescope European Coordinating Facility (ST-ECF/ESA) and the Canadian Astronomy Data Centre (CADC/NRC/CSA).

%------------------------------------------------------------------------------------------------------------------------------
%\bibliography{Library}

%----------------------------------------------------------------------------------------------------------

\end{document}